\documentclass[
 reprint,
showpacs,preprintnumbers,
 amsmath,amssymb,
 aps,
prc,
]{revtex4-1}

\usepackage{graphicx}
\usepackage{dcolumn}
\usepackage{bm}

\begin{document}

\preprint{APS/123-QED}

\title{Role of vector channel in different classes of (non) magnetized neutron stars}

\author{Luiz L. Lopes}
\email{luiz\_kiske@yahoo.com.br}
\affiliation{Centro Federal de Educa\c c\~ao  Tecnol\'ogica de
  Minas Gerais Campus VIII, CEP 37.022-560, Varginha, MG, Brasil}
\author{Debora P. Menezes}
\affiliation{%
 Departamento de Fisica, CFM - Universidade Federal de Santa Catarina;  C.P. 476, CEP 88.040-900, Florian\'opolis, SC, Brasil 
}%
102

\date{\today}

\begin{abstract}
We study how the magnetic field and non-standard vector channels affect hadronic, quark and hybrid stars. In the hadronic phase, we 
use the QHD model, with its standard $\sigma\omega\rho$ mesons, and
compare the results with the ones obtained with the inclusion of the
strangeness hidden $\phi$ meson. In the quark phase, we use the
standard SU(3) NJL and compare the results with the version that takes
into account the vector channel $G_v(\bar{\psi}\gamma^\mu\psi)$. 
Magnetic fields are taken into account via chaotic magnetic field approximation.
\end{abstract}


\maketitle

\section{Introduction \label{sec1}}

Super-powered pulsars with a magnetic field reaching as high as
$10^{15}$ G in the surface and
yet, values beyond $10^{18}$ G  in their cores (due to the scalar
Viral theorem~\cite{Shapiro}) are believed to exist in nature.
The idea of such  magnetic field several times stronger than the ones usually found in pulsars were proposed by Duncan in~\cite{Duncan1,Duncan2}
to explain two distinct objects: soft gamma-ray repeaters (SGRs) and anomalous X-ray pulsars (AXPs). Up to date, both objects are
believed to be different manifestation of a single type of compact
star: the magnetar~\cite{Sandro}. { A catalogue with updated information on
magnetars can be found in \cite{catalogue}. Presently, it contains
data on 15 SGRs and 14 AXPs.}

To describe a magnetar we revisit the theory of neutron stars, dense objects that are maintained by the equilibrium  of gravity 
and the  degeneracy pressure of the fermions. To explain the high
masses observed, besides the degeneracy pressure,  there  
must be another repulsive channel. Today it is well known that the baryon interaction is repulsive at extremely short distances. 
 In the present work, we have divided the magnetars in three different possible types.

The first one is a hadronic neutron star - or  simply neutron star for short -  described  within the Quantum hadrodynamics (QHD)
formalism~\cite{Serot}. In this model the baryon interaction is mediated by
scalar and vector mesons exchange. Within the QHD, we construct an equation of state (EoS), a relation between pressure and energy density.
With the EoS in hand, we use the  Tolman-Oppenheimer-Volkoff
(TOV) equations~\cite{TOV} to obtain the macroscopic properties of the neutron stars.
 We also allow the hyperon onset in the nuclear bulk due to the high densities reached in the
neutron star core.  The possibility of the hyperon formation is,  to
some extent, an open puzzle~\cite{Glen,Lopes2013, Fortin2017, Oertel2016,Debi2016}. 
 Recently, it was shown that the hyperon threshold can be suppressed by either very fast rotation
  \cite{Negreiros_2013} or by strong magnetic fields in a formalism that allows the coupling of
  Maxwell--Einstein equations in a self consistent way \cite{Lorene, Dex4}.
These conditions, however, are not expected in most of the observed pulsars
and the cases studied so far show that
  the results are model dependent.
We also dedicate an entire
section to describe the role of the non-standard $\phi$ meson, a repulsive
 strangeness hidden channel, which is relevant to describe massive hyperonic stars.

On the other hand, once the internal composition of the neutron stars
is not fully established, our second research considers the
possibility that the compact stars are actually quark stars, composed
of  deconfined quarks. This proposal is based on the
Bodmer-Witten conjecture~\cite{Witten,Bodmer}, which states that
strange matter is actually the ground state of  all  matter at high densities.
If this is true, the entire star is converted into quark matter in a
finite amount of time. 

Moreover, even if strange matter is not the ground
state, quarks can still be present in the core surrounded by hadronic
matter. This is the third  star type, called hybrid star. A nice
argument  to corroborate this idea is based on the large $N_c$ expansion as shown in ref.\cite{Mc}. As the quark chemical potential exceeds the constituent quark mass,
the increase of the pressure produces a phase where chiral symmetry is
restored.  With this hypothesis, for sufficiently high densities this matter becomes strange
quark matter. 

We study quark and hybrid star formalism within the three flavors
SU$(3)$ Nambu Jona Lasinio (NJL)  model~\cite{Nambu,Deb2009,DebJCAP2019,Lopes2016}, which is known
to satisfy expected QCD chiral symmetry aspects. As in the case of hadronic stars, we dedicate an entire section
to describe the role of the non-standard $G_v$  coupling, a repulsive vector channel which stiffens the EoS. 

 The role of the magnetic field on the neutron star properties is nowadays a very active field
~\cite{Dex2018,Lalit,Lee,GusakovPRD,Dex5,Dex6,Ro2017,Mauro}. In the present work, the influence of the magnetic field is studied in the context of the chaotic magnetic field
approximation~\cite{Zel,Lopes2015,china,Lopes2016}, which allows us to construct a thermodynamic consistent pressure. There is no doubt that
the ideal situation is to use the LORENE code~\cite{Lorene,Dex2018,Dex5,Dex6,Ro2017}, which performs a numerical computation of the neutron star by taking into account
Einstein-Maxwell equations and equilibrium self consistently. Unfortunately, this calculation is not always feasible for all purposes,
besides the very high computational price inherent to it. As already shown~\cite{isop}, the pressure needs to be a scalar and the chaotic magnetic field fulfills this requirement and avoids anisotropies~\cite{Lopes2015}, ~as it yields a locally effective isotropic pressure~\cite{Mauro}. Moreover, the chaotic magnetic field satisfies some constraints related to the distribution of the magnetic field in magnetar interiors, as discussed in ref.~\cite{debi2018}. Additional discussion 
about the validity of the chaotic magnetic field approximation can be found at scope and at the end of the manuscript.

\section{Magnetars as neutron stars \label{sec2}}

The natural tool to describe strong interacting matter is the quantum chromodynamics (QCD), which describes the interaction of 
quarks and gluons. However, except at a very narrow region of low density and high temperature - which turned out to be 
the opposite of what one would expect in neutron stars interiors - the
Lattice QCD (LQCD) produces no results.
To overcome this issue, we use  effective models. Here we present  the quantum hadrodynamics (QHD). The QHD considers the baryons as the fundamental
degrees of freedom and describes their interactions via mesons exchange~\cite{Serot}. 

To produce reliable neutron star properties we need to be able to reproduce realistic physical quantities that are known from 
phenomenology. There are five well known properties of symmetric
nuclear matter at the saturation point: the saturation density itself ($n_0$),
the effective nucleon mass ($M^{*}/M$), the compressibility $(K)$, the symmetry energy ($S_0$) and the binding energy per baryon ($B/A$)~\cite{Glen}.
Besides them, it is also worth noticing that the study of the slope of
the symmetry energy ($L$) - which has non-negligible implications on
the neutron star macroscopic properties  - has had a great development
in the last years, although the debate about its true value is still
going on~\cite{Rafa,Lopes2014,Tsang,
  Micaela2017,Lattimer2014,Pais2016,
Dex2019,Prov2019}. To fulfill this constraint
we use a slightly modified GM1 parametrization, which reduces the slope
from 94 MeV to 88 MeV  by reducing  $(g_{N\rho}/m_\rho)^2$ from its original value of 4.410 fm$^2$ to 3.880 fm$^2$. This modification also causes a small reduction in the symmetry energy from 32.5 MeV to  30.5 MeV.

The QHD Lagrangian in this work reads
(Eq. (\ref{EL1})):

\begin{widetext}
\begin{eqnarray}
\mathcal{L}_{QHD} = \sum_b \bar{\psi}_b \bigg [\gamma^\mu(i\partial_\mu -e_bA_\mu - g_{b,\omega}\omega_\mu  - g_{b,\rho} \frac{1}{2}\vec{\tau} \cdot \vec{\rho}_\mu)
- (m_b - g_{b,\sigma}\sigma ) \bigg ]\psi_b     + \frac{1}{2} m_v^2 \omega_\mu \omega^\mu 
   \nonumber \\ + \frac{1}{2} m_\rho^2 \vec{\rho}_\mu \cdot \vec{\rho}^{ \; \mu}   + \frac{1}{2}(\partial_\mu \sigma \partial^\mu \sigma - m_s^2\sigma^2)  
    - U(\sigma)  - \frac{1}{4}F^{\mu \nu}F_{\mu \nu}  - \frac{1}{4}\Omega^{\mu \nu}\Omega_{\mu \nu} -  \frac{1}{4}\bf{P}^{\mu \nu} \cdot \bf{P}_{\mu \nu}  , \label{EL1} 
\end{eqnarray}
\end{widetext}
in natural units. 
 $\psi_b$  are the baryonic  Dirac fields. Here, not only nucleons can be present, but we also consider the possibility of hyperon creation in neutron star core.
Because of the Pauli principle, as the number density increases, so
does the Fermi energy. Ultimately the Fermi energy of the nucleons
exceeds the mass of heavier baryons, and the conversion of some
nucleons into hyperons~\cite{Glen} become energetically favorable.  The $\sigma$, $\omega_\mu$ and $\vec{\rho}_\mu$ are the mesonic fields.  The $g's$
 are the Yukawa coupling constants that simulate the strong interaction,
 $m_b$ is the mass of the baryon $b$, $m_s$, $m_v$,  and $m_\rho$ are
 the masses of the $\sigma$, $\omega$, and $\rho$ mesons respectively,
 $e_b$ is the electric charge of the baryon $b$, $A_\mu$ is the electromagnetic
 four-potential and $\vec{\tau}$ are the Pauli matrices. 
 The antisymmetric mesonic field strength tensors are given by their
 usual expressions as presented in~\cite{Glen}, i.e.,  
$F^{\mu \nu} = (\partial^\mu A^\nu - \partial^\nu A^\mu)$,
$\Omega^{\mu \nu} = (\partial^\mu \omega^\nu - \partial^\nu \omega^\mu )$ and
$\bf{P}_{\mu \nu} = (\partial_\mu \overrightarrow{\rho}_\nu -  \partial_\nu \overrightarrow{\rho}_\mu ) - g_\rho(\overrightarrow{\rho}_\mu \times \overrightarrow{\rho}_\nu ) $. 
 The $U(\sigma)$ is the self-interaction term introduced in ref.~\cite{Boguta} to reproduce some of the saturation properties of the nuclear matter and is given by:
 \begin{equation}
U(\sigma) =  \frac{1}{3!}\kappa \sigma^3 + \frac{1}{4!}\lambda \sigma^{4} \label{EL2} .
\end{equation}

As magnetars are stable macroscopic objects, we need to describe a
neutral, chemically stable  matter and hence, leptons are added as
free Fermi gases described by
\begin{equation}
 \mathcal{L}_{lep} = \sum_l \bar{\psi}_l [i\gamma^\mu(\partial_\mu -e_lA_\mu) -m_l]\psi_l , \label{EL3}
 \end{equation}
 where the sum runs over the two lightest leptons ($e$ and $\mu$).

Let us now give the parameters of the GM1 model as well as the
prediction of the physical quantities and their inferred values from
phenomenology~\cite{Glen,Glen2, Dutra2014,Micaela2017}. In Tab. (\ref{TL1}) we resume these values.

\begin{widetext}
\begin{center}
\begin{table}[ht]
\begin{center}
\begin{tabular}{|c|c||c|c|c||c|c|}
\hline 
  & Parameters & &  Phenomenology  & GM1 & Masses ($MeV$) \\
 \hline
 $(g_{N\sigma}/m_s)^2$ & 11.785 $fm^2$ &$n_0$ ($fm^{-3}$) & 0.148 - 0.170 & 0.153 & $M_\Lambda$ = 1116\\
 \hline
  $(g_{N\omega}/m_v)^2$ & 7.148  $fm^2$ & $M^{*}/M$ & 0.7 - 0.8 & 0.7 & $M_\Sigma$ = 1193 \\
  \hline
  $(g_{N\rho}/m_\rho)^2$ & 3.880  $fm^2$ & $K$ ($MeV$)& 200 - 315&  300  & $M_\Xi$ = 1318\\
 \hline
$\kappa/M_N$ & 0.005894 & $S_0$ ($MeV$) & 30 - 34 &  30.5 & $m_e$ = 0.511  \\
\hline
$\lambda$ &  -0.006426 & $B/A$ ($MeV$) & 15.7 - 16.5 & 16.3 & $m_\mu$ = 105.6\\
\hline 
$M_N$ &  939 $MeV$ & $L$ ($MeV$) & 36 - 113 & 88 & - \\ 
\hline
\end{tabular}
 
\caption{GM1 model  parameters and physical quantities infered from experiments
  \cite{Glen2, Dutra2014, Micaela2017}. } 
\label{TL1}
\end{center}
\end{table}
\end{center}
\end{widetext}

As we allow the hyperon onset we also have to fix the hyperon-mesons coupling constants. Unlike the nuclear matter, we have very little experimental 
information about hyperonic matter. The main term is the hyperon potential depth fixed at the saturation density. However, just the $\Lambda$ hyperon
has the potential depth well fixed at -28 $MeV$~\cite{Glen2}. The knowledge of the other potential depths are known with a lower degree of precision. Unfortunately 
the knowledge of the hyperon potential depth is not enough to fix all  constants, once different sets of coupling constants reproduce
the same potential values. Even worst is the fact that these different sets of the coupling constants, yet predicting the same potential depth,
cause  large variations on the neutron star properties~\cite{Glen}.  So, in order to reduce the large number of free parameters we fix the well known $\Lambda$ 
potential depth $U_\Lambda$ = -28 $MeV$, and use symmetry group theory to fix all the other hyperon-meson coupling constants. To be more specific, we 
use the hybrid SU(6) group~\cite{Pais} to fix all the vector mesons and a nearly SU(6) group to fix the scalar ones as presented in \cite{Lopes2013}.
 All the hyperon-mesons coupling constants are presented bellow in Eq. (\ref{EL4}).

\begin{eqnarray}
\frac{g_{\Lambda\omega}}{g_{N\omega}} = \frac{g_{\Sigma\omega}}{g_{N\omega}} = 0.667,  \quad \frac{g_{\Xi\omega}}{g_{N\omega}} =  0.333, \label{EL4} \\
\frac{g_{\Sigma\rho}}{g_{N\rho}} = 2.0 \quad \frac{g_{\Xi\rho}}{g_{N\rho}} = 1.0 , \quad \frac{g_{\Lambda\rho}}{g_{N\rho}} = 0.0,  \nonumber \\ 
\frac{g_{\Lambda\sigma}}{g_{N\sigma}} = 0.610 , \quad \frac{g_{\Sigma\sigma}}{g_{N\sigma}} =  0.396 , \quad \frac{g_{\Xi\sigma}}{g_{N\sigma}} = 0.113 .\nonumber
\end{eqnarray}

To solve the equations of motion, we use the mean field approximation (MFA), where the meson fields are replaced by their expectation values, i.e:  $\sigma$ $\to$ $\left < \sigma \right >$ = $\sigma_0$,   $\omega^\mu$ $\to$ $\delta_{0 \mu}\left <\omega^\mu  \right >$ = $\omega_{0}$  and   $\rho^\mu$ $\to$ $\delta_{0 \mu}\left <\rho^\mu  \right >$ = $\rho_{0}$.
Applying the Euler-Lagrange formulation to Eq.~(\ref{EL1}) we obtain,
for a baryon in the absence of electric field, the following equation of motion:

\begin{equation}
[\gamma_0(i\partial^0 - g_{b, \omega}\omega_0 - g_{b, \rho}\rho_0) - \gamma_j(i\partial^j -e_bA^j) - M^{*}_b]\Psi = 0, \label{EL5}
\end{equation}
where we define $M^{*}_b~\doteq~m_b - g_{b,\sigma}\sigma_0$ as the effective baryon mass.

For an uncharged baryon, $e_bA^j$ is always zero. Using the quantization rules ($E = i\partial^0$, $i\partial^j = k$) we easily obtain the eigenvalue for the
energy:  

\begin{equation}
E_b = \sqrt{k^2 + M^{*2}_b} + g_{b,\omega}\omega_0 + g_{B,\rho} \frac{\tau_3}{2}  \rho_0 .  \label{EL6}
\end{equation}

In the case of magnetars, as expected, the magnetic field plays a crucial role. So, to study its effect, we start with a static external magnetic field
in the $z$ direction (these conditions will be relaxed latter). To accomplish that we fix: $A_2 = A_3 = 0; A_1 = -B_0y$.
This choice results in a quantum harmonic oscillator like equation,
whose  eigenvalue solution is well known ~\cite{Lopes2012,Pal,Peng, Rhabi}:
\begin{equation}
E_b = \sqrt{k_z^2 + M^{*2}_b +2\nu|e|B_0} + g_{b,\omega}\omega_0 + g_{B,\rho} \frac{\tau_3}{2}  \rho_0 , \label{EL7}
\end{equation}
where $\nu$ is a discrete parameter called Landau level (LL). For the
leptons, since they are all charged and  are not affected by the strong force, we have:
\begin{equation}
E_l = \sqrt{k_z^2 + m^{2}_l +2\nu|e|B_0}. \label{EL8}
\end{equation}

To construct the equation of state (EoS) for this many body system of
leptons and strongly interacting baryons we use the Fermi-Dirac statistics.
 As the thermal energy of a stable neutron star is much lower
  than the Fermi energy of its particles, $T= 0$ is a good
  approximation.  For the uncharged baryons
 the distribution is isotropic and the solution for the energy density is straightforward~\cite{Greiner2}:

\begin{equation}
\epsilon = \frac{1}{\pi^2}\int_0^{k_f} \sqrt{k^2 + M^{*}_b}k^2 dk . \label{EL9}
\end{equation}

For the charged fermions (baryons and leptons), not only the energy spectrum is quantized but also the distribution in the plane perpendicular to the
magnetic field. In this case the energy density reads~\cite{Lopes2012,Peng}:

\begin{equation}
\epsilon =  \frac{|e|B_0}{2\pi^2} \sum_{\nu} \eta(\nu) \int_0^{k_f} \sqrt{k_z^{2} + M^{*}_b + 2\nu|e|B_0} dk_z, \label{EL10}
\end{equation}
where $k_f$ is the Fermi momenta of the particle and $\eta(\nu)$ is the degeneracy  of the LL $\nu$. The first Landau Level is non-degenerate and the others are
two-fold degenerate. The sum over the LL ends at the closest integer
at which the square of the Fermi momenta of the particle is still positive.

In MFA the contribution of the mesonic fields to the energy density  is given by~\cite{Glen,Lopes2012,Lopes2013}
\begin{equation}
\epsilon_m =  \frac{1}{2}\bigg ( m_s^2\sigma_0^2 + m_v^2\omega_0^2  + m_\rho^2\rho_0^2 \bigg ) + U(\sigma) , \label{EL11}
\end{equation}

The total energy density is the sum of the energy density of all fields (baryons, leptons and mesons).
 Finally the expected values of the mesonic fields are calculated
either from the Euler-Lagrange equations or by imposing that the total energy density be
stationary at fixed baryon density~\cite{Glen}:

\begin{equation}
\bigg ( \frac{\partial \epsilon}{\partial \sigma_0} \bigg ) =  \bigg ( \frac{\partial \epsilon}{\partial \omega_0} \bigg)
= \bigg ( \frac{\partial \epsilon}{\partial \rho_0} \bigg ) = 0 . \label{EL12}
\end{equation}

To calculate every particle population at a fixed density we impose
electric charge neutrality and chemical equilibrium:

\begin{equation}
\mu_{bi} = \mu_n -e_{bi}\mu_e , \mu_\mu =  \mu_e ; \quad \sum_f e_fn_f = 0 , \label{EL13}
\end{equation} 
where $\mu_{bi}$ and $e_{bi}$  are the chemical potential and electric
charge of the i-th baryon respectively. At zero temperature,
the chemical potentials coincide with the energy eigenvalues given in Eqs.~(\ref{EL6}) and~(\ref{EL7}); $\mu_e$ and $\mu_\mu$ are the electron
and muon chemical potential respectively; $n$ is the number density
and the sum in $f$ runs over the two fermions.

Now, to construct the EoS we calculate the pressure via thermodynamics:
\begin{equation}
p = \sum_f \mu_f n_f - \epsilon , \label{EL14}
\end{equation}
where the sum runs over all fermions.

\subsection{Contribution of the magnetic field itself}

Now let's take a closer look at the magnetic field itself. As we known from Einstein's general relativity,
every form of energy should be taken into account via the energy-momentum tensor. As the magnetic field itself
has energy it will ultimately contribute in the TOV
equations~\cite{TOV}. Although its contribution to the energy 
is simple, the contribution to the pressure is not trivial. The reason
is that for a constant magnetic field in the 
$z$ direction, the stress tensor is not diagonal, but reads $B = diag(B^2/2,B^2/2, -B^2/2)$~\cite{Gravitation}.

The straightforward way to solve this problem is a fully relativistic  numerical integration
of Einstein's field equations. However it demands a very high computational price. Once we are
in the realm of effective fields, we can look for options. The
standard approach in the literature~\cite{Lopes2012,Ryu,Panda,Mallick,Dex1,Casali,Dex2,Dex3} is
to consider  only the perpendicular component of the pressure.
In this case, the contribution of the pressure is equal to the
contribution of the energy, $p = \epsilon = B^2/2)$. However, this
approach cannot be justified by first principles, overestimates the influence of the magnetic field
 and is in disagreement with more precise calculations~\cite{Lorene, debi2018}.
Another common approach is to split the pressure into two parts, the
so called perpendicular and parallel pressures and then solve an
axisymmetric version of the TOV equations~\cite{Mallick2, Zubairi, Paret}. However
this approach has a big issue once the thermodynamical concept of 
pressure cannot depend on the direction because it is a scalar. Also,
it was already discussed in the literature that the pressure
is ultimately isotropic even in the presence of a magnetic
field~\cite{isop}.
Besides the problem of the thermodynamical consistency in the presence
of a magnetic field, there is also a problem regarding the stability of the magnetic field itself. As discussed in
the literature, an uniform magnetic field in the $z$ direction is
unstable~\cite{insta1,insta2}. A way to avoid both problems lies in the
concept of chaotic magnetic field, as originally introduced by Zeldovich~\cite{Zel} in the 60's:  {\it ``It is possible to describe the effect of the magnetic field by
 using the pressure concept only when we are dealing with a small-scale chaotic field (pag. 158)"}.
So we relax the condition of an uniform magnetic field in the $z$
direction and assume the chaotic magnetic 
field, whose stress tensor reads: $diag(B^2/6, B^2/6, B^2/6)$, thus avoiding the anisotropy problem and yielding
$p = \epsilon/3$, a radiation pressure formalism. Within this approach
the contribution of the magnetic field 
to the EoS reads~\cite{Lopes2015,china}:
\begin{equation}
\epsilon = \epsilon_M + \frac{B_0^2}{2} \quad ; \quad p = p_M + \frac{B_0^2}{6} , \label {EL15}
\end{equation}
where $M$ stands for the matter.

We now go back to the issue of an uniform magnetic field. The more powerful magnetars are expected to have 
a magnetic field around $10^{15}G$ in their surface~\cite{Duncan1, Duncan2}. Although fields of these magnitudes do not
affect the main properties of neutron stars, fields larger than $10^{18}G$ are expected in the magnetar core
due to the scalar Virial theorem~\cite{Shapiro}. To simulate the magnetic field growing towards the core 
we use an energy density dependent approach, as introduced in~\cite{Lopes2015}:

\begin{equation}
B(\epsilon) = B_0 \bigg ( \frac{\epsilon_M}{\epsilon_c} \bigg ) ^\gamma + B_{surf} , \label{EL16}
\end{equation}
where $\epsilon_M$ is the matter energy density and $\epsilon_c$ is
the energy density at the core for zero magnetic field configuration,
$B_{surf}$ is the magnetic field at the surface of the magnetar, for which we use $10^{15}G$, $B_0$ is the
maximum magnetic field expected in the core, which we consider to be
of the order of 3 $\times$ $10^{18}G$ to account for the scalar Virial theorem, and
$\gamma$ is the only free parameter of our model. To study its
influence on the neutron star properties, 
we use $\gamma$ = 4 and $\gamma$ = 6.    Now $B(\epsilon)$ replaces
$B_0$ in Eq.~(\ref{EL15}), only in the therm of self-energy - $B^2/2$ and  $B^2/6$ - , but not in the 
matter term. We make this choice for two reasons: assure that the 
energy eigenvalue is correct, and the fact that using a variable
magnetic field also in the matter does not significantly affect the
results as showed in ref.~\cite{Casali}. 
Concerning this approach, a word
of caution is important: a density dependent magnetic field violates Maxwell equations
as discussed in \cite{alloy} and a rearrangement term, never
calculated, would be necessary. { Moreover, as the chaotic magnetic field is spherically symmetric, neutron star deformations are beyond the scope of this work. To study deformations and 
deviations from the spherical symmetry, one needs to use poloidal magnetic fields as presented in ref.~\cite{Dex5,Ro2017}, or even more complex geometries as a combination between toroidal and poloidal fields.}

\begin{figure*}[ht]
\begin{tabular}{cc}
\includegraphics[width=5.6cm,height=6.2cm,angle=270]{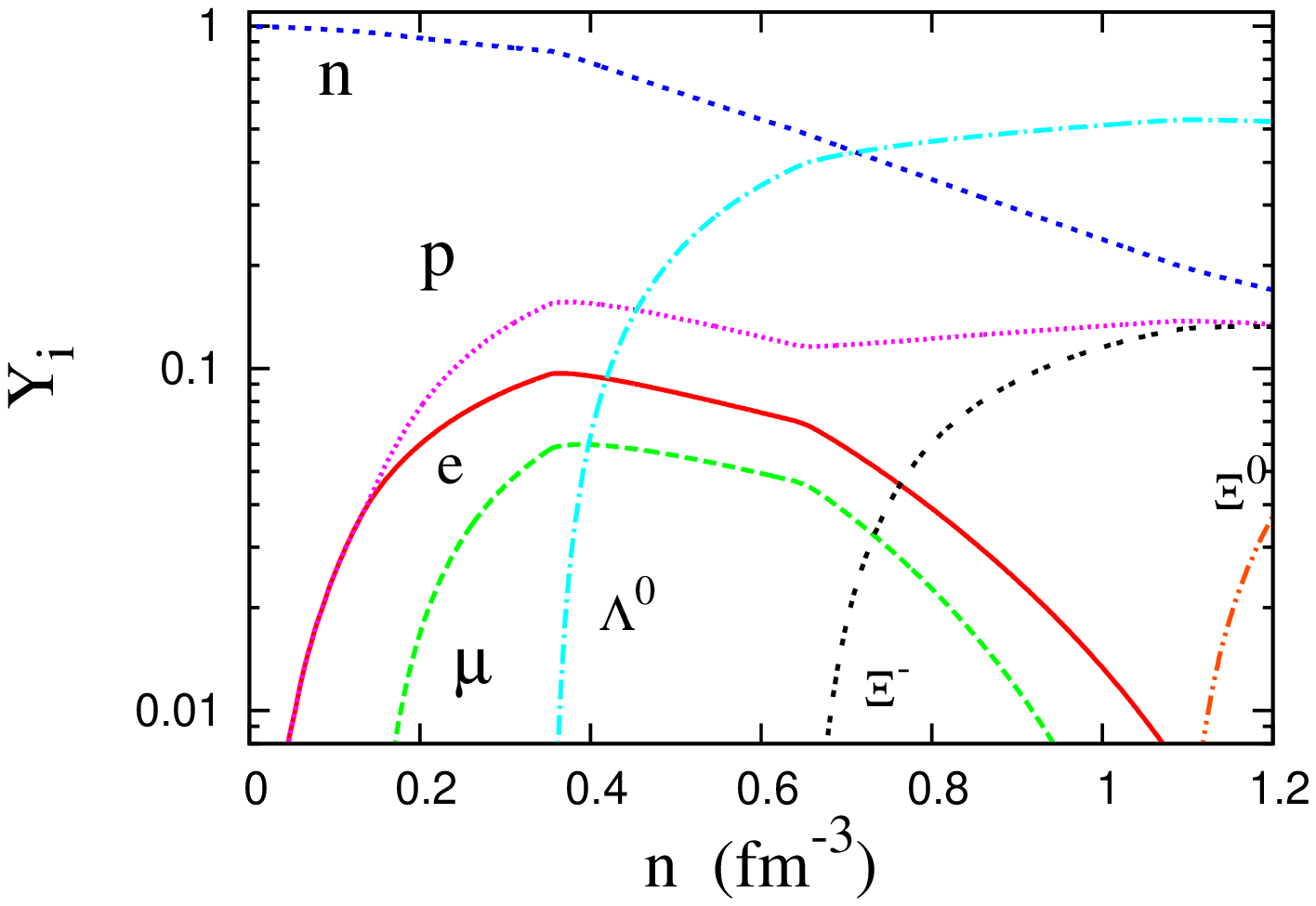} &
\includegraphics[width=5.6cm,height=6.2cm,angle=270]{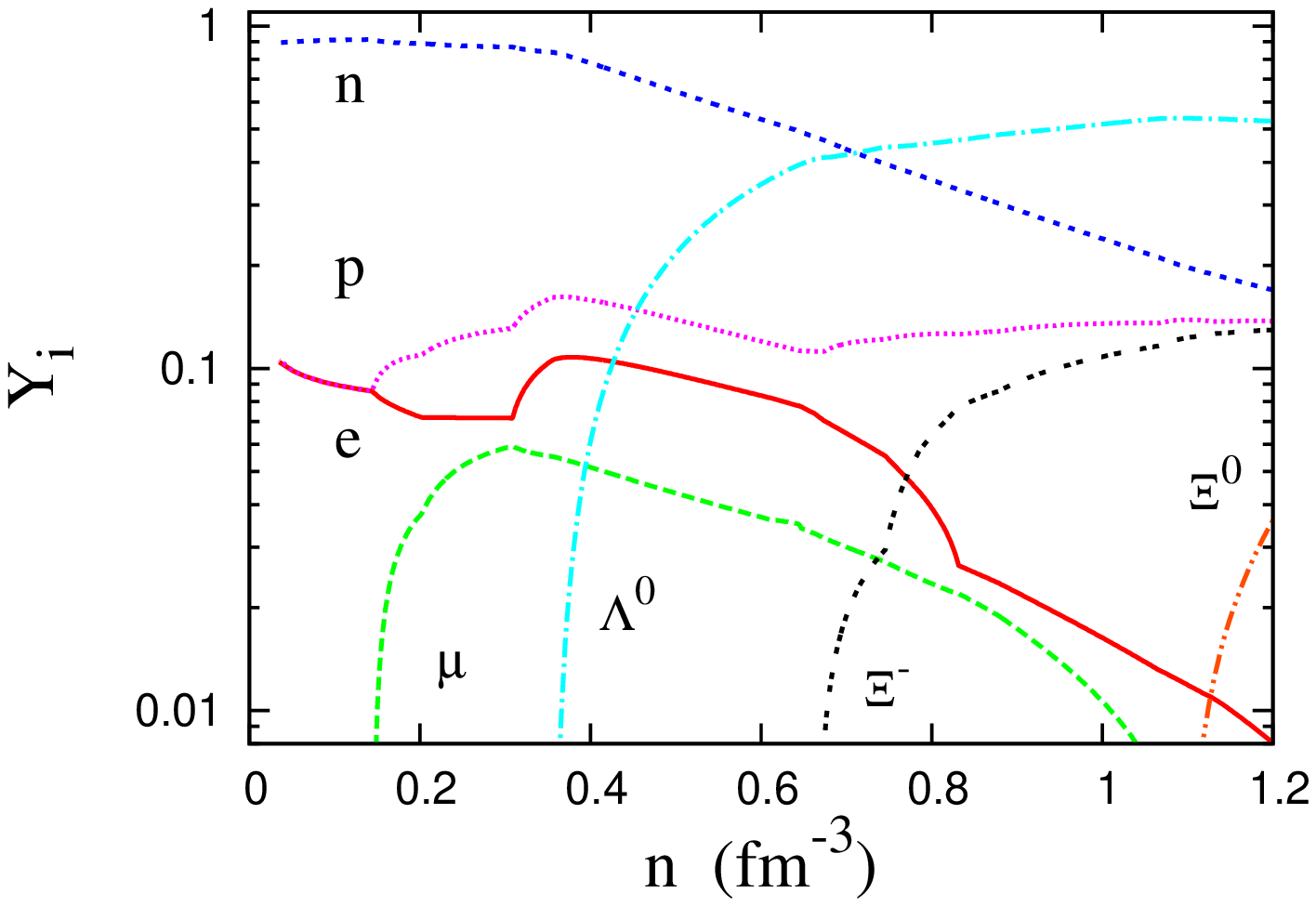} \\
\end{tabular}
\caption{(Color online) Particle population for zero magnetic field
  (right) and for $B_0$ = 3 $\times 10^{18}G$ (left). The Landau
quantization induces several kinks and descontinuities at small densities. } \label{FL1}
\end{figure*}

\begin{figure*}[ht]
\begin{tabular}{cc}
\includegraphics[width=5.6cm,height=6.2cm,angle=270]{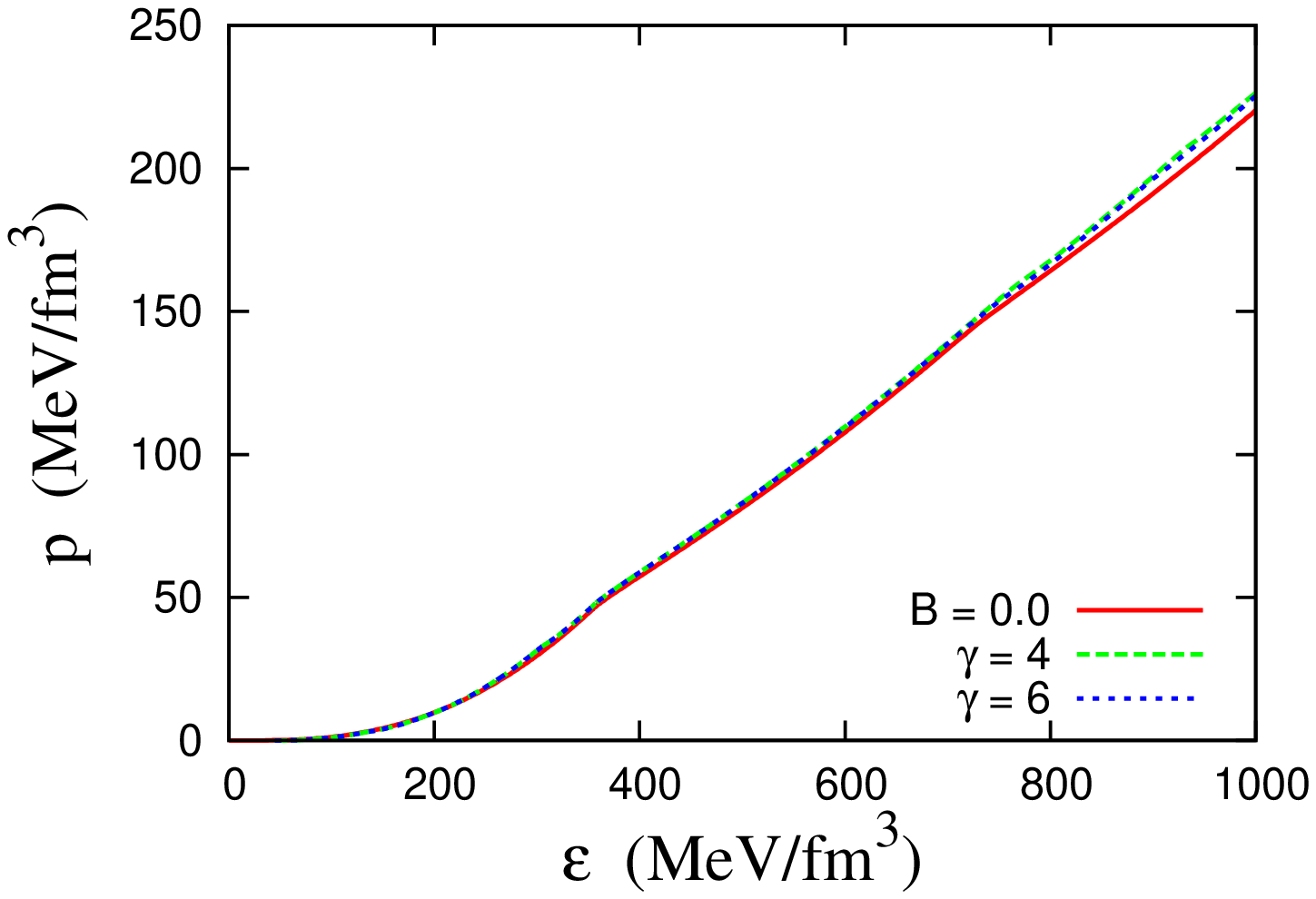} &
\includegraphics[width=5.6cm,height=6.2cm,angle=270]{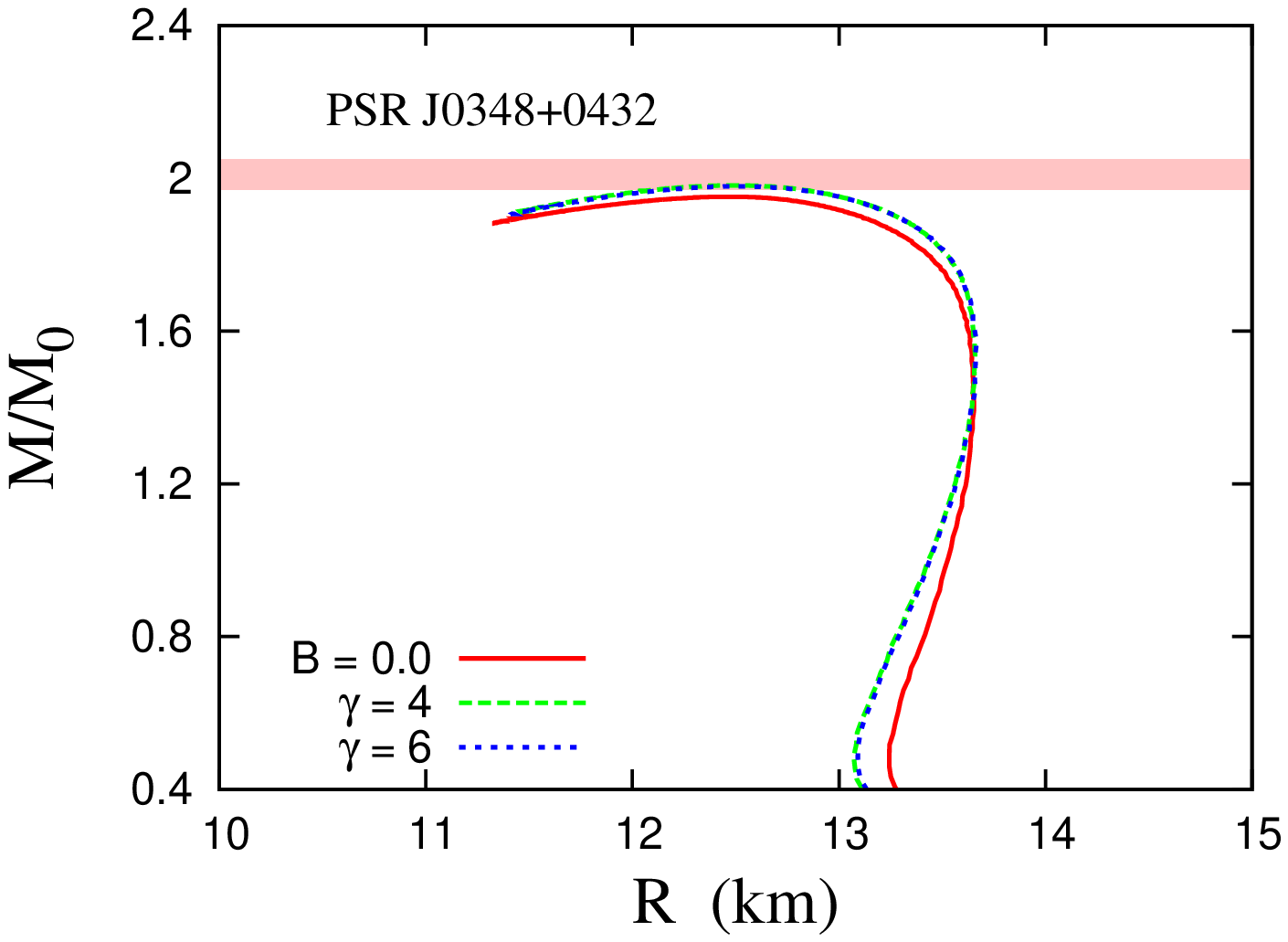} \\
\end{tabular}
\caption{(Color online) (right) EoS and (left) mass-radius relation
  for a neutron star with zero magnetic field and two variations of
  density dependent magnetic fields. Within the chaotic magnetic field
  approximation the increase of the maximum mass is very small and the
  results are independent of the free parameters.} \label{FL2}
\end{figure*}

\subsection{Results}

We plot in Fig.~(\ref{FL1}) the particle population for zero magnetic field and $B_0 = 3 \times 10^{18}G$.
The hyperon population is strong dependent on the hyperon-meson coupling constants. With our choice that 
the couplings are constrained to the SU(6) symmetry group, we see that the $\Lambda^0$ is the first hyperon 
to appear around $n = 0.35 fm^{-3}$.  Also, with this approach, the
$\Sigma$ triplet is suppressed,  the $\Xi^{-}$ 
appears around 0.66 $fm^{-3}$ while the onset of the $\Xi^{0}$ occurs
at densities beyond those found in the neutron star core. When the
magnetic field is present, two main differences arise in comparison
with the results in non magnetized matter: the first one is that the 
charged particle population is favored at low densities since they are coupled to the magnetic field.
The second is the presence of many kinks, especially at low densities due the dependence with the discrete LL.
At low densities there are few LL available. As the density increases, the number of LL  grows and 
approaches the continuum and the particle population is similar to
those found in the absence of the magnetic field.

For the macroscopic properties of the neutron stars, we plot in Fig.~(\ref{FL2}) the EoS and mass-radius relation for zero magnetic field and a magnetic field of
$3 \times 10^{18}G$ within two different variations, with $\gamma$ = 4 and 
$\gamma$ = 6. In the absence of magnetic field, the maximum neutron
star mass is 1.95 $M_\odot$ which is 
slightly below the limit of the PSR J0348+0432 in ref.~\cite{Antoniadis}. 
When the magnetic field is taken into account, the maximum mass
increases to 1.98 $M_\odot$ which agrees with the measurement of PSR J0348+0432. Also, this small increase of the mass is in agreement with more complex models in different approaches as can be seen int~\cite{Lorene,Mallick2,Dex4,Ro2017}. Moreover, comparing the results for
$\gamma$ = 4 and $\gamma$ = 6 we see that 
with the chaotic magnetic field,  the variation of the magnetic field has very little
influence on the macroscopic properties of the magnetars, a very
reasonable result, once the $\gamma$ is a non-observable parameter. 
 At the end of 2019 an even more massive pulsar was discovered, the MSP J0740+6620, which has a mass of $2.14^{+0.10}_{-0.09}$ at 68$\%$ credibility interval and $2.14^{+0.20}_{-0.18}$ at 95$\%$  credibility interval~\cite{Cromartie}. Although there is a small chance that its mass lies below 2 $M_\odot$, if it turns out to be above 2.1 $M_\odot$, a new formalism may be necessary to explain such massive pulsar~\cite{Clesio}. 

 Recent estimates of neutron star radii~\cite{Hebeler,Lattimer2013} point out that the canonical star (1.4 $M_\odot$) has a radius not much larger than 13 km. Although these results refer to non-magnetized neutron stars, we see that in our case, stars below 1.5 $M_\odot$ bear radii that are smaller than non-magnetized ones. These stars have a relatively low magnetic field; for instance, in our model a neutron star with a mass of 1.2 to 1.3 solar masses has a central magnetic field below $10^{17}~G$~\cite{Lopes2015}. However, as can be { seen} from Fig.~(\ref{FL2}), stars with masses above 1.6 solar masses have, in fact, larger radii than non-magnetized ones. For a 1.8 $M_\odot$ neutron star, its central magnetic field lies around 2-5 $\times 10^{17}~G$. The maximum magnetic field reached in the core of the maximum mass neutron star is also very close to the value of $B_0$, but never surpasses it. The main macroscopic properties of the maximum mass neutron stars, as well the radii of canonical 1.4 $M_\odot$ stars are plotted in Tab.~(\ref{TL2}).

  \begin{table}[ht]
\begin{tabular}{|c|c|c|c|c|c|}
\hline
 $\gamma$ & $M/M_\odot$ & $ R (km)$ & $\epsilon_c$ ($MeV/fm^3$) & $B_c$ ($ 10^{18}G)$ & $R_{1.4} (km)$  \\
\hline
 $B=0$ & 1.95 & 12.47   & 976 & 0.0  & 13.63  \\
\hline
 4         & 1.98 & 12.47  & 970 & 2.98  & 13.62 \\
 \hline
6          & 1.98 & 12.51  & 963 & 2.96 & 13.58  \\
\hline
\end{tabular}
 \caption{Neutron stars main properties for zero magnetic field and two variations of
  density dependent magnetic fields, both with $B_0 = 3 \times 10^{18}G$.}\label{TL2}
 \end{table}

\subsection{The importance of the vector channel}

The GM1 parametrization, which is used in this work was fitted to
describe the properties of nuclear matter at the saturation point. To
describe the hyperon threshold in the neutron star core, a new set of
hyperon-meson couplings was necessary. Due to our choice of  the SU(6)
symmetry group, only one free parameter was needed, which was fixed by
the imposition that $U_\Lambda$ = -28 $MeV$. However, when hyperons are
present not only new coupling constants arise, 
but also a new vector meson is possible, the strangeness-hidden $\phi$ meson, whose Lagrangian
reads~\cite{Lopes2013,Lopes2018,kapusta}:

\begin{figure*}[ht]
\begin{tabular}{cc}
\includegraphics[width=5.6cm,height=6.2cm,angle=270]{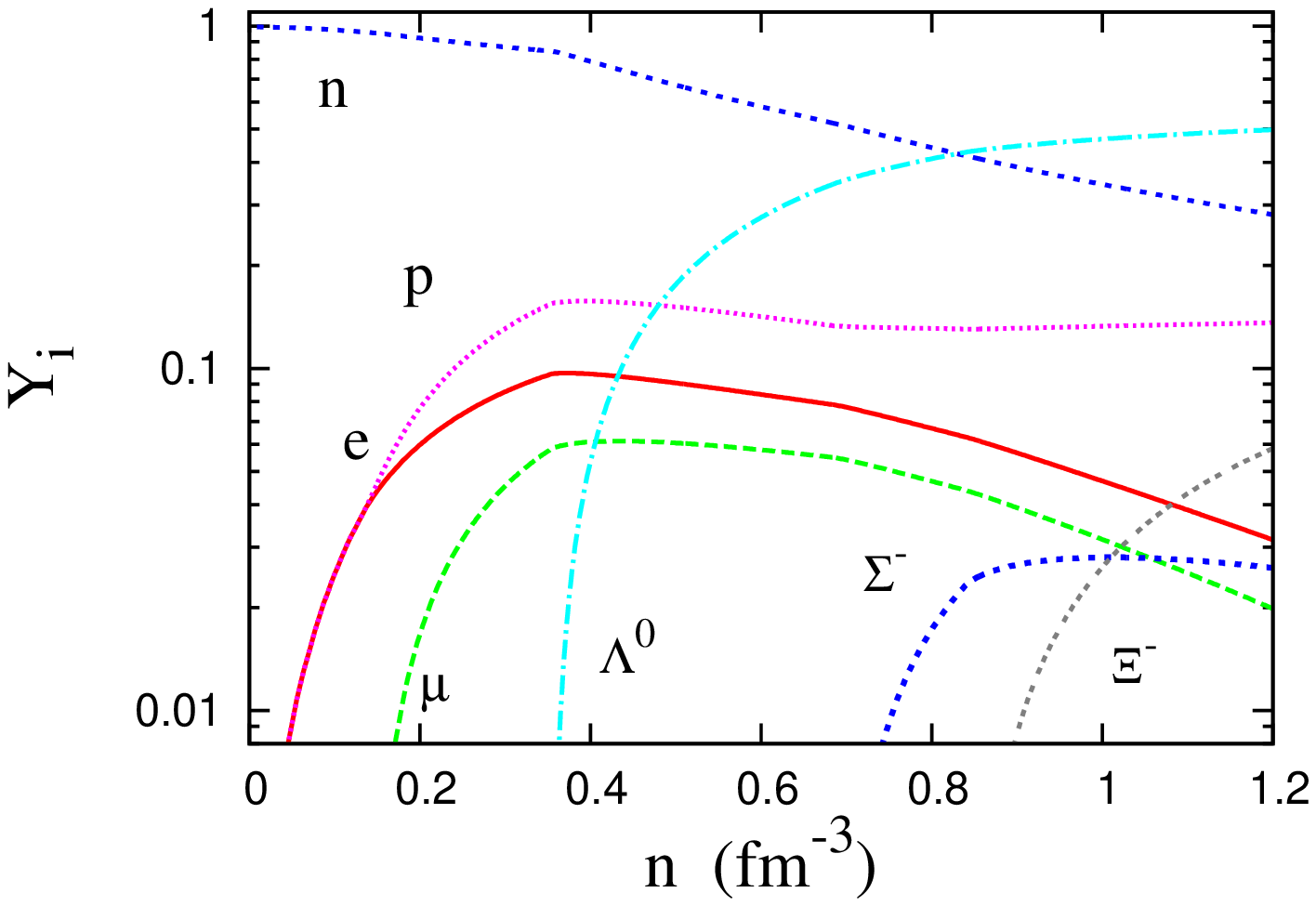} &
\includegraphics[width=5.6cm,height=6.2cm,angle=270]{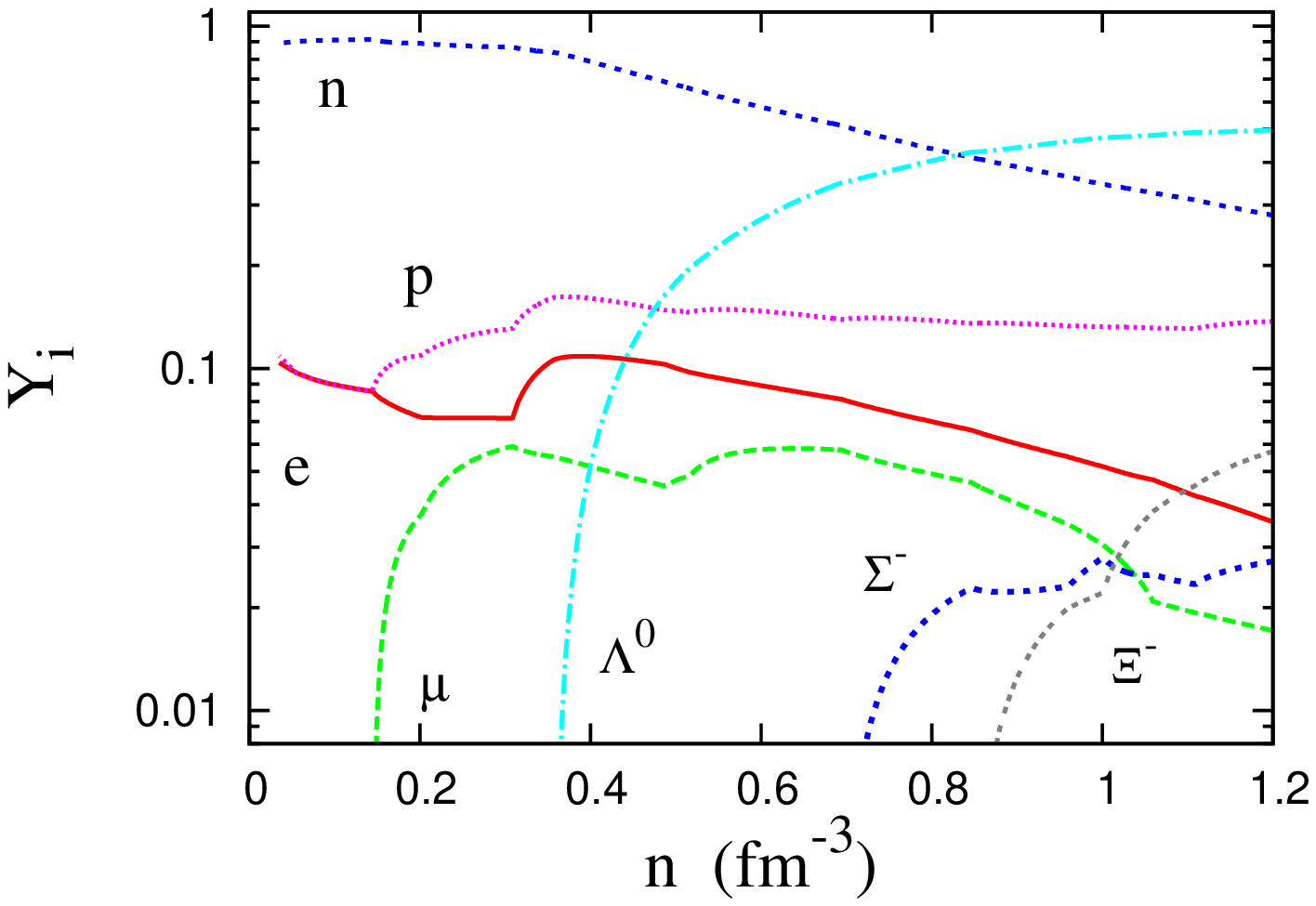} \\
\end{tabular}
\caption{(Color online) Particle population for zero magnetic field and for $B_0$ = 3 $\times 10^{18}G.$ The $\phi$ meson causes
the $\Sigma^{-}$ be present  and pushes the $\Xi$'s to very high densities. } \label{FL3}
\end{figure*}

\begin{figure*}[ht]
\begin{tabular}{cc}
\includegraphics[width=5.6cm,height=6.2cm,angle=270]{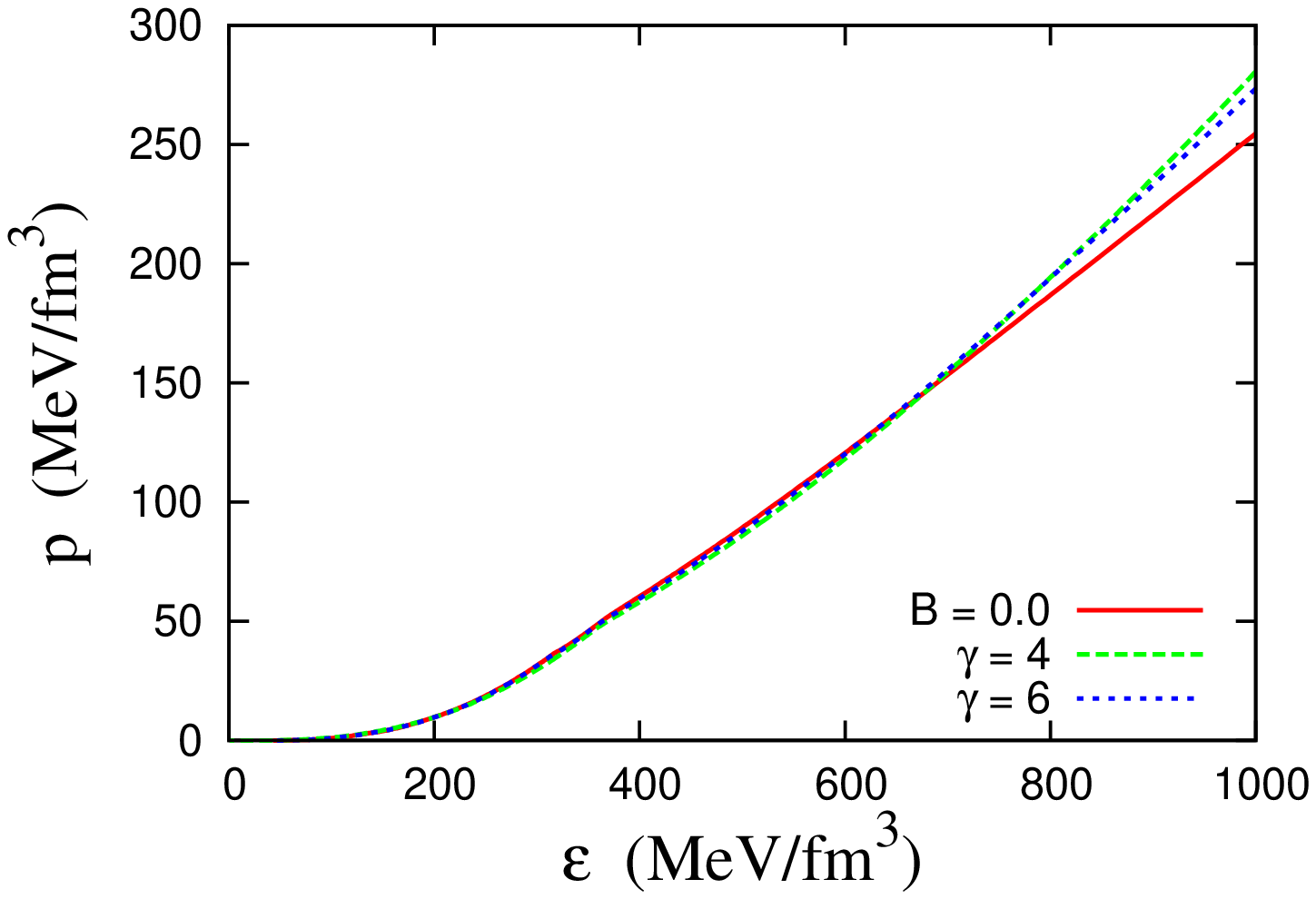} &
\includegraphics[width=5.6cm,height=6.2cm,angle=270]{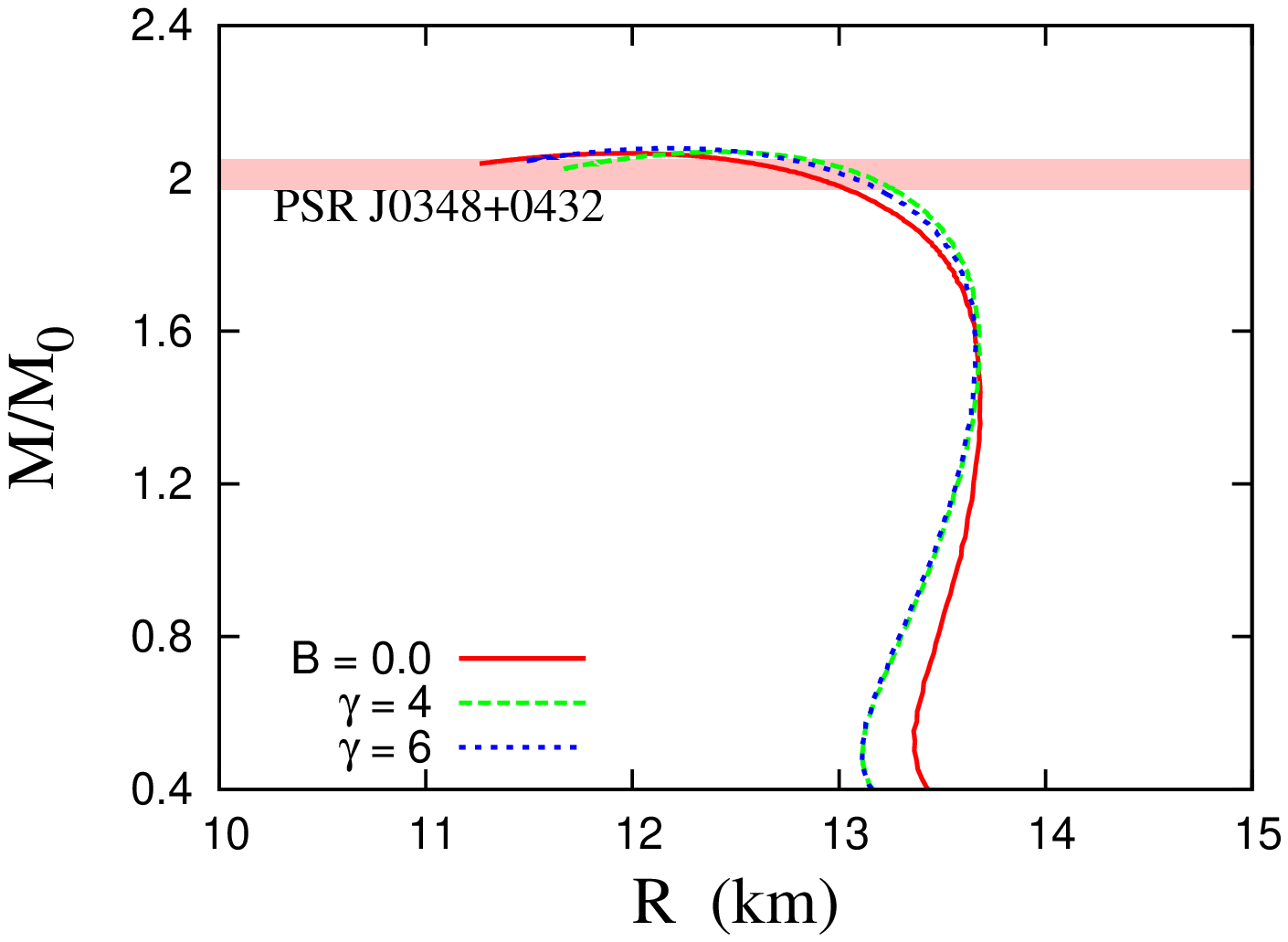} \\
\end{tabular}
\caption{(Color online) EoS and Mass-radius relation for a neutron
  star with zero magnetic field and two 
variations of magnetars with the $\phi$ meson.
All neutron stars results now agree with the PSR J0348+0432. } \label{FL4}
\end{figure*}

\begin{equation}
\mathcal{L} = g_{Y,\phi}\bar{\psi}_Y(\gamma^\mu\phi_\mu)\psi_Y + \frac{1}{2}m_\phi^2\phi_\mu\phi^\mu - \frac{1}{4}\Phi^{\mu\nu}\Phi_{\mu\nu} , \label{EL17}
\end{equation}

The $\phi$ meson is analogous to the $\omega$ meson and its expected
value is also calculated via MFA~\cite{Lopes2013,kapusta}. As it is a vector meson,
the $\phi$ field increases the repulsion between the hadrons and
consequently increases the maximum neutron star mass. Nevertheless, adding
a new meson implies on a new set of free parameters in the
Lagrangian. However, using the SU(6) symmetry group, all
hadron-$\phi$ couplings are already constrained and within this approach, no new
free parameter is needed at all. From the symmetry group we have~\cite{Lopes2013}:

\begin{eqnarray}
\frac{g_{\Lambda\phi}}{g_{N\omega}} = \frac{g_{\Sigma\phi}}{g_{N\omega}} = -0.471,  \quad \frac{g_{\Xi\phi}}{g_{N\omega}} =  -0.943, \label{EL18} \\
 \quad \frac{g_{N\phi}}{g_{N\omega}} = 0.0.  \nonumber 
\end{eqnarray}

Now in Fig.~(\ref{FL3}) we plot the particle population for $\beta$-stable matter with the $\phi$ meson 
for zero magnetic field and for a magnetic field equal to 3 $\times$
$10^{18}$ G. For low densities, as the $\phi$ meson does not couple to
the nucleon, there is no difference when compared to results without it. Once the $\Lambda$ appears,
 the $\phi$ field is no longer null and a new repulsion channel arises. As can be seen
from Eq.~(\ref{EL18}), the repulsion of the $\Xi$ hyperons is
stronger, which causes a suppression of these particles allowing
the onset of the $\Sigma^{-}$ hyperon, which was absent when the
$\phi$ meson was not taken into account. As in the last section, when a strong magnetic field is present, the particle densities are quantized due the LL and many kinks appear.

Now in Fig.~(\ref{FL4}) we plot the EoS and the mass-radius relation
for neutron stars and magnetars with the $\phi$ meson.  As the $\phi$
meson increases the hyperon chemical potentials, it reduces the
strangeness fraction of the hyperons in neutron stars core. Moreover, 
the $\phi$ meson itself  enters in the pressure calculation, causing
the EoS to become stiffer. All these effects combined
cause a significant stiffening and an increase in the maximum mass of
non-magnetized neutron stars from  1.95 $M_\odot$
to 2.06$M_\odot$, which puts the neutron stars in agreement with the PSR J0348+0432~\cite{Antoniadis}  and even with MSP J0740+6620~\cite{Cromartie}. It also causes an increase of the radii for stars above 1.9$M_\odot$, as they are near the maximum mass when the $\phi$ meson is absent, but not so near when it is included.
As far as magnetars are concerned, 
when the magnetic field is present the maximum mass has a small rises from 2.06 $M_\odot$ to 2.07 $M_\odot$ and 2.08 $M_\odot$,
depending of the value of $\gamma$ This shows that the chaotic magnetic field is a free-parameter way to introduce magnetic effects 
on the macroscopic properties of neutron stars.These small increase of the maximum mass are again in agreement of more complex
calculations~\cite{Dex3,Lorene}. We resume the main properties of the
neutron stars and magnetars with the $\phi$
 meson in Tab.~(\ref{TL3}).

\begin{table}[ht]
\begin{tabular}{|c|c|c|c|c|c|}
\hline
 $\gamma$ & $M/M_\odot$ & $ R (km)$ & $\epsilon_c$ ($MeV/fm^3$) & $B_c$ ($ 10^{18}G)$ & $R_{1.4} (km)$  \\
\hline
 B=0 & 2.06 & 11.96   & 1122 & 0.0  & 13.63  \\
\hline
 4         & 2.07 & 12.19  &  998 & 2.66 & 13.62    \\
 \hline
6          & 2.08 & 12.18  & 1060 & 2.83 & 13.58  \\
\hline
\end{tabular}
 \caption{Neutron stars main properties for zero magnetic field and
two variations of density dependent magnetic fields, both with $B_0 = 3 \times 10^{18}G$ with
the $\phi$ meson take into account. }\label{TL3}
 \end{table}

To summarize this section, let's say that we have studied the theory of magnetars as a hadronic neutron star, allowing the hyperon onset due the
high densities reached in the core. Using the GM1 model for the nuclear matter and the symmetry group arguments to fix the hyperon-meson coupling constant we show that non-magnetized neutron stars can reach a maximum mass
of 1.95$M_\odot$, which is subtly below the constraint of the PSR J0348+0432. This issue is solved when we 
consider the effects of magnetic field or the strangeness-hidden $\phi$ meson, which does not couple to the nucleon,
 and rises the maximum mass
to 2.06$M_\odot$. The influence of the magnetic field is studied using the chaotic field approximation, and 
its effect is to always increase the maximum mass. Without the $\phi$ meson this increase is  about $2\%$
and with it, the increase drops to only $1\%$. In all cases, the
influence of the magnetic field seems independent
of the non-observable parameter $\gamma$.

\section{Magnetars as quark stars}

As stated out earlier, the true ground state of the hadronic matter
 might not be composed  of protons and neutrons but rather, of strange quark matter, as 
predicted by Bodmer and Witten~\cite{Bodmer,Witten}. If this is true,
pulsars with central densities above a certain limit are quark stars,
or strange stars. The main difference of quark stars to the conventional neutron stars, composed 
of baryons, is the fact that while the neutron star is bounded by gravity, the quark star is self-bounded, i.e, is the
strong force that keeps matter cohesive against the degeneracy pressure of the quarks.

\begin{widetext}
\begin{center}
\begin{table}[ht]
\begin{center}
\begin{tabular}{|c|c||c|c|c||c|}
\hline 
  & Parameters & &  Phenomenology  & SU(3) NLJ \\
 \hline
 $m_u = m_d$ & 5.5 $MeV$  & $m_\pi$ ($MeV$) & 128 -138 & 138 \\
 \hline
  $m_s$ & 135.7 $MeV$   & $m_\eta$ ($MeV$) & 487  & 549  \\
  \hline
  $\Lambda$ & 631.4 $MeV$   & $m_\sigma$ ($MeV$) & 668 &  700  \\
 \hline
$G_s\Lambda^2$ & 1.835 & $f_\pi$ ($MeV$) & 93 &  93   \\
\hline
$K\Lambda^5$ &  9.29 & $f_\eta$ ($MeV$) & 94.3 & 84 - 102 \\
\hline 
\end{tabular}
 
\caption{SU(3) NJL  parameters and physical quantities infered from experiments
\cite{Kun2}. } 
\label{TL4}
\end{center}
\end{table}
\end{center}
\end{widetext}

To describe a quark star we need to use a quark matter EoS. Again, the official tool to describe quarks
via standard model is the QCD and once more we need an effective model. As the quark matter is expected to 
restore the chiral symmetry at high densities, we seek a model with this feature. To accomplish this task we use the
SU(3) version of the NJL model~\cite{Nambu,DebPRC2009,DebPRC2014}.  The SU(3) NJL Lagrangian includes a scalar, 
a pseudo-scalar and the t'Hooft six-fermion interaction - needed to
model the axial symmetry breaking~\cite{Kun1,Kun2}. Its Lagrangian density reads:

\begin{widetext}
\begin{eqnarray}
\mathcal{L}_{NJL} =  \bar{\psi}_f  [\gamma^\mu(i\partial_\mu -e_fA_\mu) -  m_f]\psi_f   - \frac{1}{4}F_{\mu\nu}F^{\mu\nu} + \nonumber \\
+ G_s \sum_{a=0}^8[(\bar{\psi}\lambda_a\psi)^2 + (\bar{\psi}\gamma_5\lambda_a\psi)^2] 
- K\{det[\bar{\psi}(1 + \gamma_5)\psi + det[\bar{\psi}(1 -\gamma_5)\psi]\} 
\end{eqnarray}
\end{widetext}
where $\psi_f$ are the quark Dirac fields, with three flavors, $m_f$ = diag$(m_u, m_d, m_s)$ are the
current quark masses, $e_f$ represents the quark electric charge, $\lambda_a$ are the eight Gell-Mann flavor matrices and $G_s$ and $K$ 
are coupling constants. Unlike the QHD model for baryons,  where the interaction is mediated by massive mesons,
 the NJL model has no mediator, and the interaction is a direct quark-quark point-like scheme (see ref.~\cite{DebPRC2009,Kun2} to see the Feynman diagrams).
This makes the NJL  a non-renormalizable model, and a cutoff is needed to obtain physical results. The SU(3) NJL predicts five main physical parameters:
the $\pi$, $\eta$ and $\sigma$ meson masses, as well  pion and $\eta$ decay coupling constants, $f_\pi$ and $f_\eta$. The parameters,
physical predictions and experimental values are given in Tab. (\ref{TL4}). 

Assuming mean field approximation (MFA) we can rewrite the quark-quark interaction in terms of the scalar condensates ($(\bar{\psi}\psi)2$ = 
$2\langle\bar{\psi}\psi\rangle - \langle\bar{\psi}\psi\rangle^2)$~\cite{MB}.
To construct the EoS for quark matter, we need to construct the thermodynamical potential, once p = $-\Omega$. For each flavor, 
the pressure has three components, the in medium, the vacuum and the magnetic contribution ($p_i$ = $p^{med}_i +p^{vac}_i + p^{mag}_i$),
 besides the condensates ($\phi_f = \langle\bar{\psi}_f\psi_f\rangle$)
 that also contribute to the pressure. Following ref.~\cite{Kun2,DebPRC2009,DebPRC2014,MB} and assuming T = 0, we have:

\begin{equation}
\Omega = -\sum_f p_f + 2G_s(\phi_u^2 + \phi_d^2 +\phi_s^2) - 4K\phi_u\phi_d\phi_s ,
\end{equation}
where the contributions are:

\begin{eqnarray}
p_f^{med} =  \frac{N_c|e_f|B}{2\pi^2} \sum_\nu \eta(\nu)\int_{0}^{k_f} dk \frac{k^2}{E_f} \nonumber \\
p_f^{vac} = \frac{N_c}{\pi^2} \int_0^{\Lambda} dk \epsilon_v k^2  \\
p_f^{mag} = \frac{N_c(|e_f|B)^2}{2\pi^2}  \bigg [ \zeta '(-1, x_f) + \nonumber \\
- \frac{(x_f^2 - x_f)}{2}\ln x_f + \frac{x_f^2}{4} \bigg ] \nonumber ,
\end{eqnarray}
where $N_c$ is the number of colors, 
$E_f$ = $\sqrt{M_f^2 + k_z^2  +2|e_f|B}$ is the energy eigenvalue  for fermions under the
 influence of external magnetic fields and
$M_f$ are the dynamical quark masses, which depend on the condensates: 
\begin{equation}
M_f =  m_f - 4G_s\phi_i + 2K\phi_j\phi_k. \label{EG}
\end{equation}
The $\epsilon_v = \sqrt{k^2 + M^2}$ can be  seen 
as a vacuum energy eigenvalue, as it is very similar to the energy of
free fermions; 
$\zeta '$ is the derivative of the Riemann-Hurwitz zeta function
and $x_f = M_f^2/(2|e_f|B)$. For more details on the formalism, please
see ref.~\cite{DebPRC2009,DebPRC2014,MB}. 

To finish, we need to calculate the value of the condensates.  In the
same way as the pressure,  the condensates also have contributions
from medium, vacuum, and the magnetic 
field ($\phi_f = \phi_f^{med} + \phi_f^{vac} + \phi_f^{mag})$. These values are:

\begin{eqnarray}
\phi_f^{med} =  \frac{N_c|e_f|BM_f}{2\pi^2} \sum_\nu \eta(\nu) \ln\bigg ( \frac{E_f + k_f}{\sqrt{M_f^2 + 2|e_f|B}} \bigg ) \nonumber \\
\phi_f^{vac} = -\frac{N_cM_f}{2\pi^2} \bigg [ \Lambda\epsilon_v - \frac{M_f^2}{2}\ln \bigg (\frac{\epsilon_v + \Lambda}{M^2} \bigg ) \bigg ] \nonumber \\  \\
\phi_f^{mag} = -\frac{M_fN_c(|e_f|B)^2}{2\pi^2}  \bigg [ \ln\Gamma(x_f) - \frac{1}{2} \ln(2\pi)  +  \nonumber \\
+ x_f -\frac{(2x_f - 1)}{2} \ln x_f \bigg ]  \nonumber. 
\end{eqnarray}

As the dynamical masses $M_f$ depend of the condensate values and
vice-versa, these equations are solved in a self-consistently
way. Equations (\ref{EG}) are called gap equations. As in the hadronic case, leptons are added as a free Fermi gas, as required by  charge neutrality and
chemical stability.   The relations
between the chemical potentials and the number density of different particles are given by~\cite{Rhabi}:

  \begin{eqnarray}
  \mu_s =\mu_d = \mu_u + \mu_e , \quad \mbox{and} \quad \mu_e = \mu_\mu , \nonumber \\
  n_s + n_\mu = \frac{1}{3}(2n_u -n_d - n_s). \label{NEQ}
  \end{eqnarray}

The energy density is obtained
via the thermodynamic relation given in (Eq.~\ref{EL14}).

\subsection{Contribution of the magnetic field itself}

As in the case of neutron stars, we construct quark stars via the
chaotic magnetic field approximation, which gives us an isotropic contribution 
to the pressure avoiding thermodynamic issues. The energy density   and pressure are given by:

\begin{equation}
\epsilon = \epsilon_M + \frac{B_0^2}{2} \quad ; \quad p = p_M + \frac{B_0^2}{6} .
\end{equation}

We continue to  couple the density depended magnetic field to
the EoS instead   of coupling it to the number density, as presented in Eq.~(\ref{EL16}).
Additional discussion about the  validity of this approach can be found at the end of the manuscript.

\begin{figure*}[ht]
\begin{tabular}{cc}
\includegraphics[width=5.6cm,height=6.2cm,angle=270]{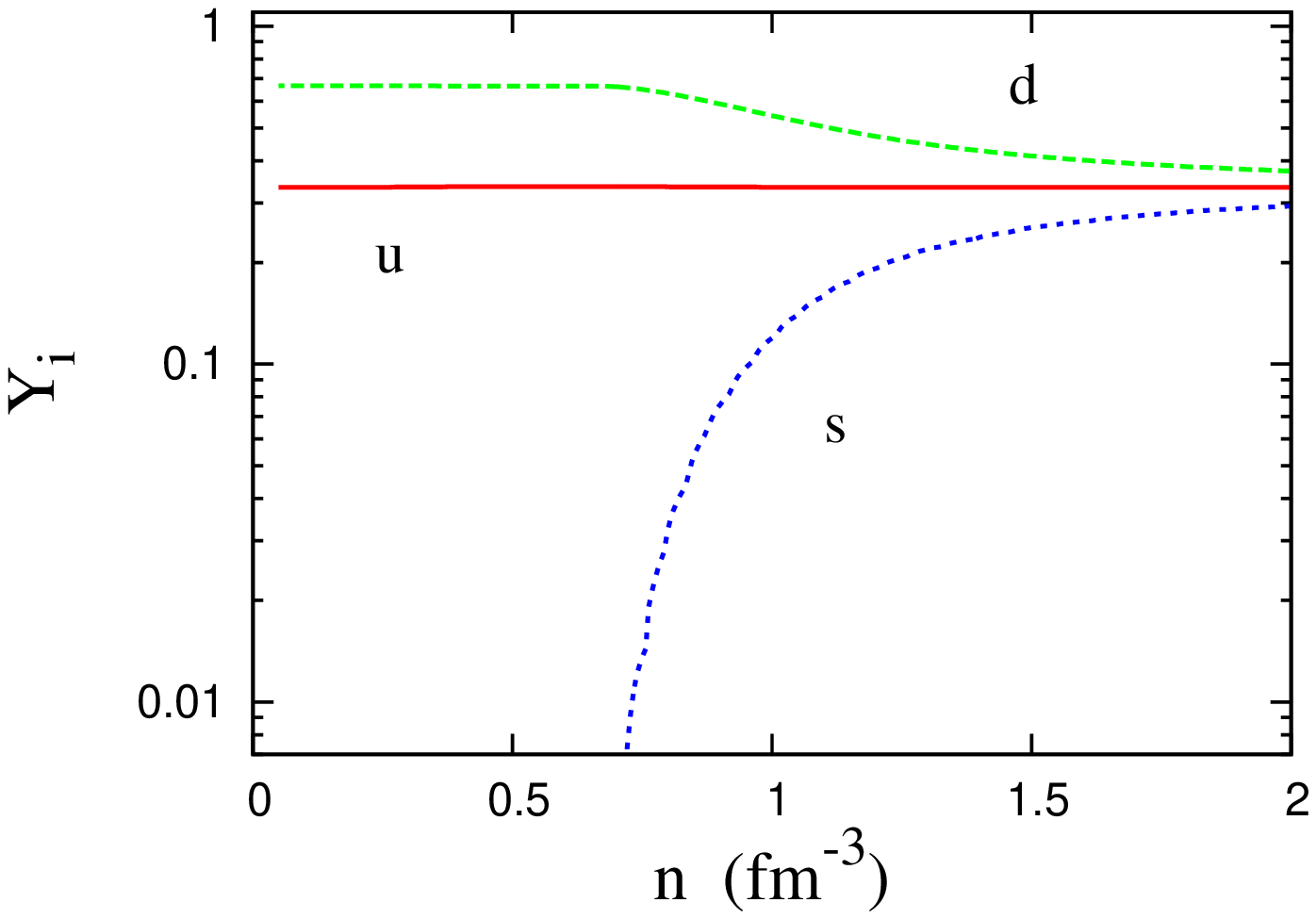} &
\includegraphics[width=5.6cm,height=6.2cm,angle=270]{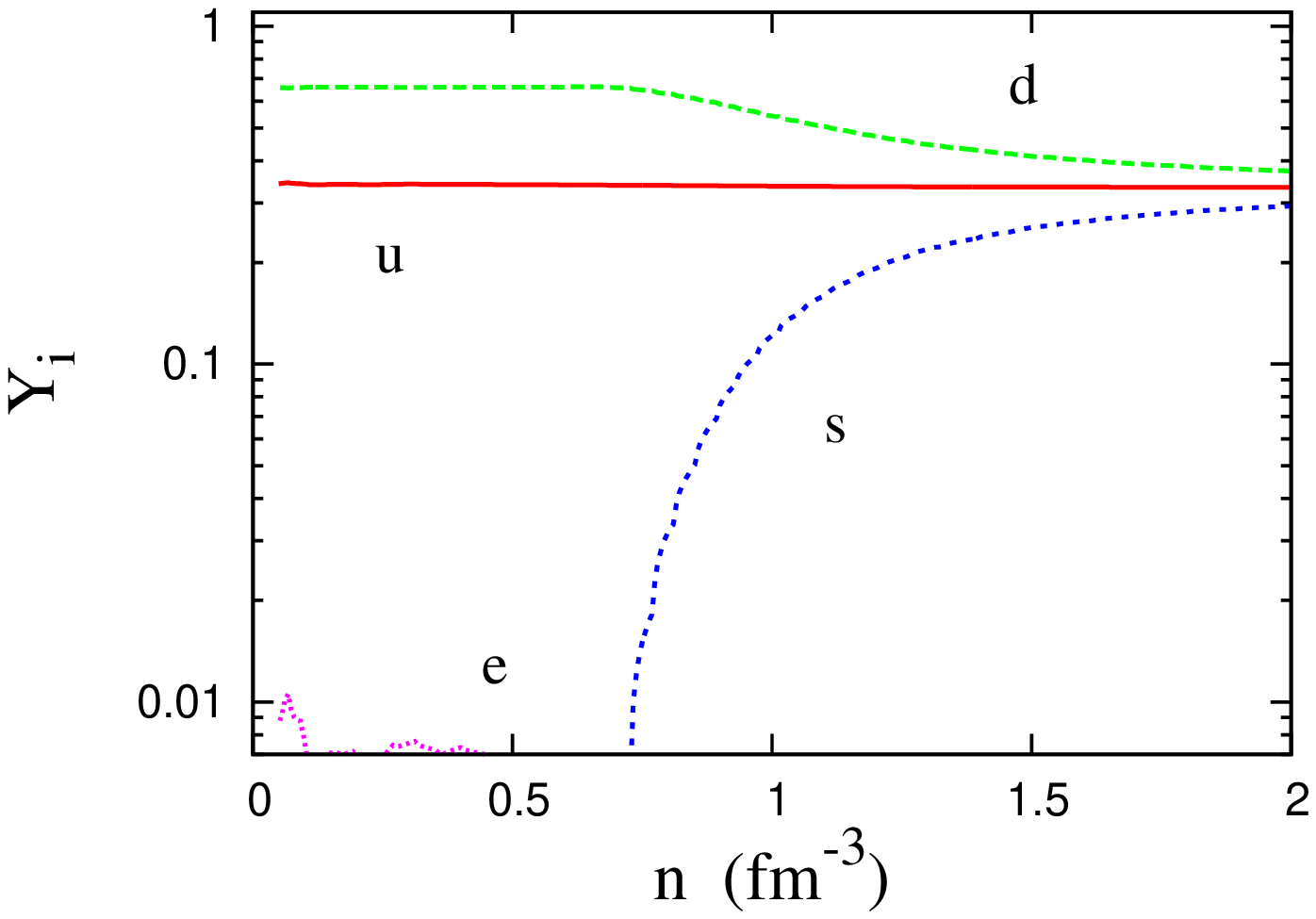} \\
\end{tabular}
\caption{(Color online) Particle population for zero magnetic field
  (left) and for $B_0$ = 3 $\times 10^{18}G$ (right). The electron population is only
significant in the presence of the magnetic field. } \label{FL5}
\end{figure*}

\begin{figure*}[ht]
\begin{tabular}{cc}
\includegraphics[width=5.6cm,height=6.2cm,angle=270]{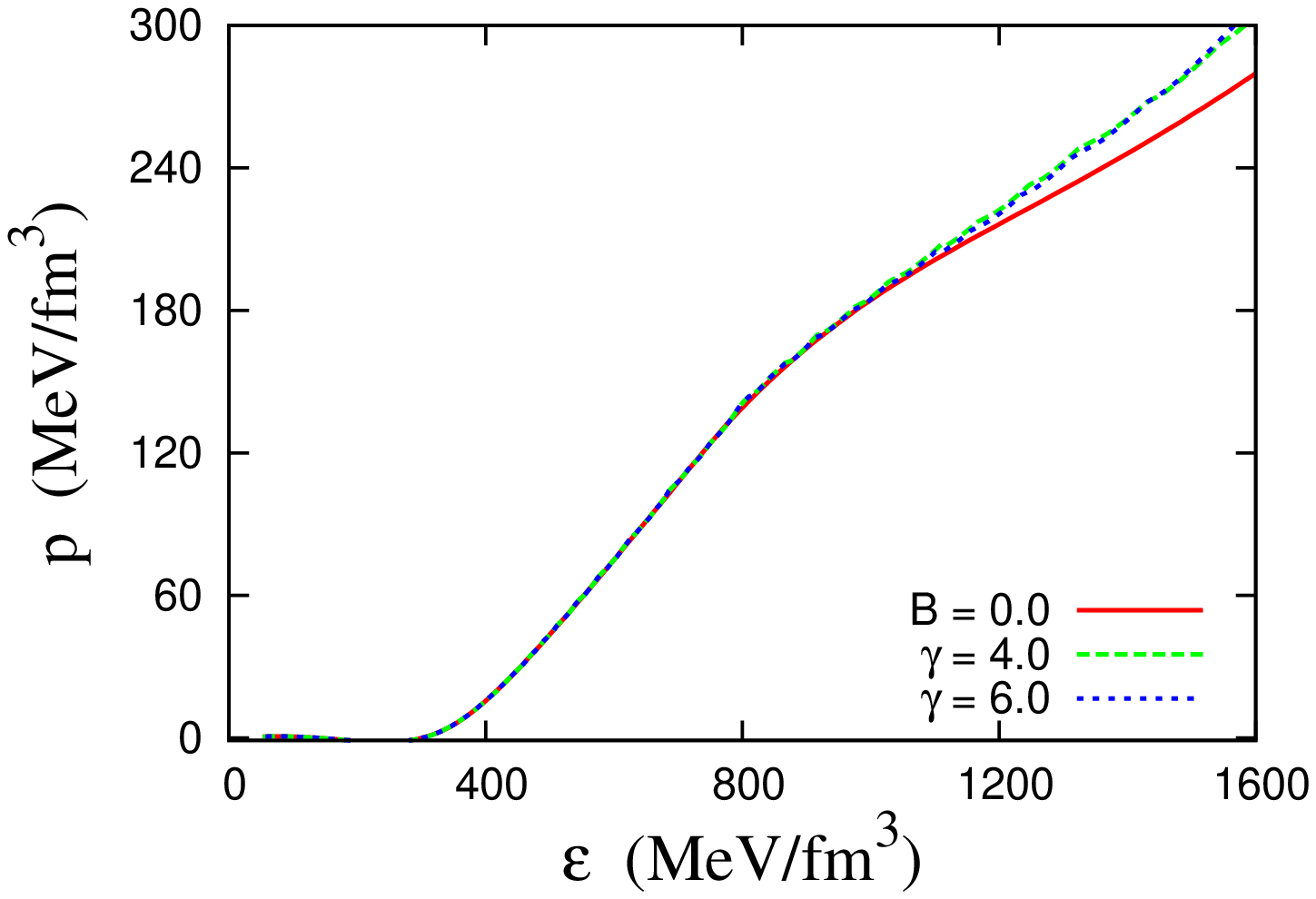} &
\includegraphics[width=5.6cm,height=6.2cm,angle=270]{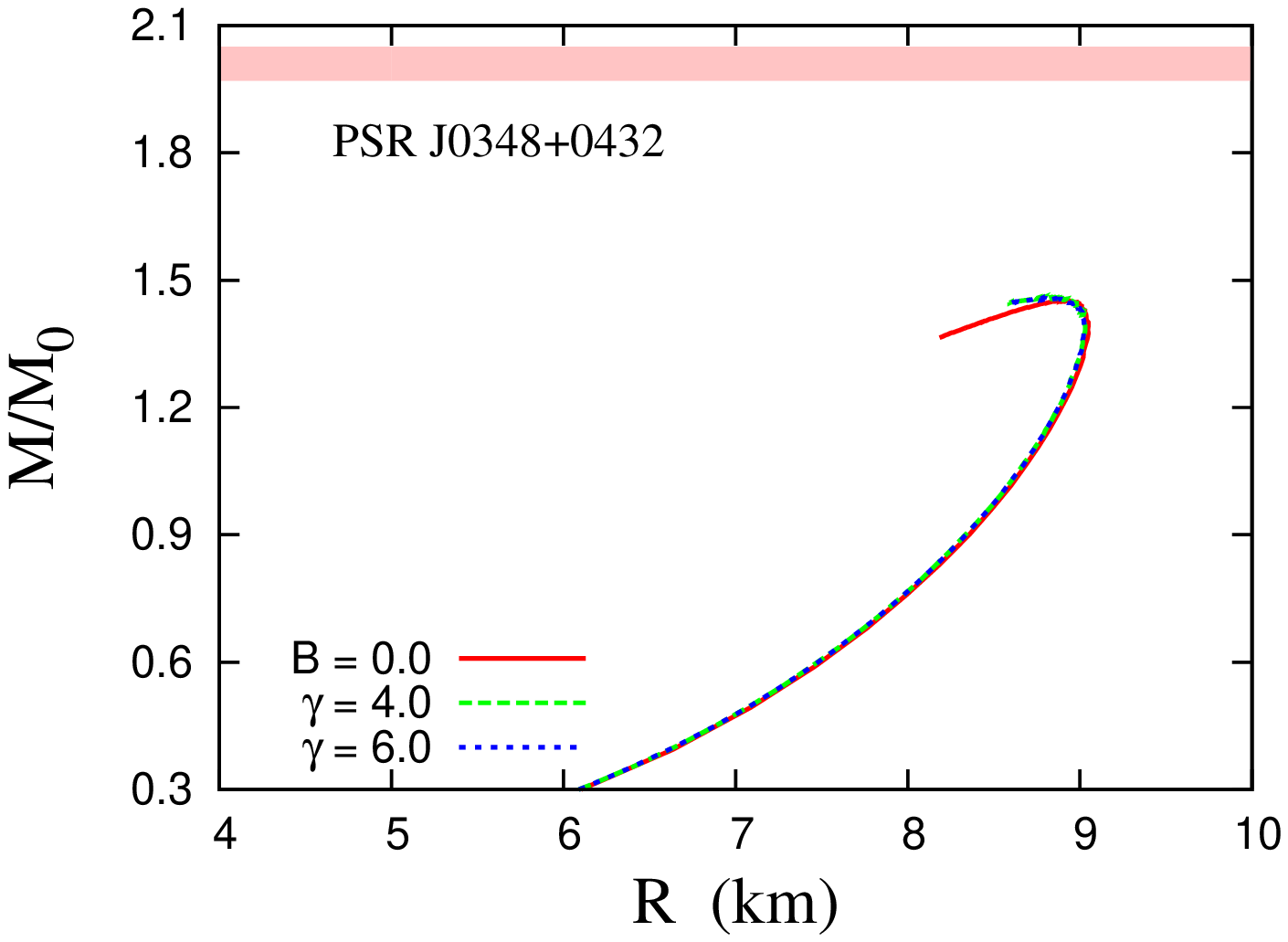} \\
\end{tabular}
\caption{(Color online) (right) EoS and (left) mass-radius relation
  for a quark star with zero magnetic field and two variations of
  density dependent magnetic fields. Within the chaotic magnetic field
  approximation the increase of the maximum mass is very small and the
  results do not dependent on the free parameters.} \label{FL6}
\end{figure*}

\subsection{Results}

In  Fig.~(\ref{FL5}) we plot the  particle population for zero magnetic field and $B_0 = 3 \times 10^{18}G$.
Unlike the baryon population, the quark population  does not
  depend strongly on  coupling constants. As  $G_s$ and $K$  coupling
constants are equal to all quarks, the main  difference is
  generated by the mass of the $s$ quark.  
With our choice following
ref.~\cite{Kun1}, the $s$ quark onset happens around 0.66 $fm^{-3}$. On other hand, we see that for zero magnetic field, the electron population is very low, reaching a maximum
of only $Y_e =0.002$ next to the $s$ quark threshold. The muons are
absent due the fact that its mass is bigger than the mass of light quarks.  
The main difference in particle population when the magnetic field is present, is the fact that the electron population is much bigger at 
low densities, reaching a maximum of $Y_e = 0.01$. Although this is
still a low value, it is five times bigger than  the one obtained
in the absence of the 
magnetic field. This is  due to the fact that the particle population couples to the
magnetic field and the electric charge. As the electron charge is
three times the value of the quark charge, its population rises,
particularly at low densities when there are 
less Landau levels. 
Even at low densities the kinks are much less evident than in hadronic
matter,  since in the last case we have the continuum Fermi sea of neutrons (uncharged)
 competing with the discrete Landau levels of protons and electrons (charged).
On the other hand, in quark matter all particles are charged.

As for the EoS itself, it is plotted in Fig.~(\ref{FL6}) alongside
some of the corresponding macroscopic properties of the quark stars, i.e.,
the mass-radius relation for zero magnetic field and for a magnetic field of
$3 \times 10^{18}G$ within two different variations of  the
  density dependence, i.e., $\gamma$ = 4 and 
$\gamma$ = 6. In the absence of the magnetic field, the maximum quark
star mass is only 1.46 $M_\odot$ which is below the limit of the PSR
J0348+0432 in ref.~\cite{Antoniadis},  ruling out  the NJL as a
model that describes massive pulsars.  Another important aspect   refers to the fact that the NJL model does not produce absolutely   stable matter at zero temperature, i.e., it does not satisfy the   Bodmer-Witten conjecture \cite{buballa_99}. However, the effects of a magnetic field  of the order of $10^{18}~G$ and a small increase of temperature \cite{veronica_2013} seem to be enough to
guarantee that the quark matter acquires stability.

When the magnetic field is taken into account, the maximum mass
increases only to 1.47 $M_\odot$. As in the case of hadronic neutron stars, this small increase of the mass is in agreement
with more complex models in different approaches as can be seen in ref.~\cite{Lorene,Mallick2}. Moreover, comparing the results for
$\gamma$ = 4 and $\gamma$ = 6 we see that  with the chaotic magnetic field,  the variation of the magnetic field has again very little
influence on the macroscopic properties of the magnetars. Also,
the effect of the magnetic field besides slightly increasing the maximum mass, causes 
a reduction on the radii of low mass quark stars. This same result was observed in hadronic neutron stars,
and shows that the chaotic field approximation is consistent with both
types of star.
The central density of a quark star is around 1500 $MeV/fm^3$, more than $50\%$ higher
than the central density of the neutron stars. Yet, the central magnetic field
is lower when compared with the hadronic case. In Tab.~(\ref{TL5}) we
resume these results,  which can be compared with the ones
  presented in Tab.~(\ref{TL2}).

  \begin{table}[ht]
\begin{tabular}{|c|c|c|c|c|c|}
\hline
 $\gamma$ & $M/M_\odot$ & $ R (km)$ & $\epsilon_c$ ($MeV/fm^3$) & $B_c$ ($ 10^{18}G)$ & $R_{1.4} (km)$  \\
\hline
 $B=0$ & 1.46 & 8.92   & 1480 & 0.0  & 9.03  \\
\hline
 4         & 1.47 & 8.85  & 1516 & 2.19 & 9.03    \\
 \hline
6          & 1.47 & 8.83  & 1534 & 2.38 & 9.02  \\
\hline
\end{tabular}
 \caption{Quark stars main properties for zero magnetic field and two variations of   density dependent magnetic fields, both with $B_0 = 3 \times 10^{18}G$.}\label{TL5}
 \end{table}

\subsection{The importance of the vector channel}

The  SU(3) NJL model parameters were fixed in order to determine the five
properties presented in Tab.~\ref{TL4}. However, additional terms can be 
added in a covariant way in the NJL model. One of these terms is a 
vector interaction, in opposition to the scalar interaction. Its
Lagrangian reads:

\begin{equation}
\mathcal{L}_{NJLv}= -G_v(\bar{\psi}\gamma^\mu\psi)^2 .\label{ELQv}
\end{equation}

In the
phase diagram, the vector term weakens and delays the phase transition of the
chiral restoration, and potentially alter the nature from
chiral transition  to the  color-superconducting (CSC) phase~\cite{kit2002}. The mathematical formalism of the
vector term shows that it acts like the $\omega$ meson in
 quantum hadrodynamical  models, creating an additional 
repulsion between the quarks and stiffens the EoS \cite{klahn}. This effect is desirable
once we need to construct an EoS stiff enough to simulate the two solar
mass PSR J0348+0432 pulsar.

When the vector channel is present an additional term is added in
the pressure (as well in the energy density via Eq.(\ref{EL14})).
In MFA this term is~\cite{Kun2}:

\begin{equation}
p =  G_vn^2 ,
\end{equation}
As its value
is zero at zero potential chemical, there is no vacuum correction
or gap equation.
 Notice that another prescription for the vector chanel in the NJL
  model is also possible. For a discussion on the different
  possibilities and their consequences, please see ref.~\cite{DebPRC2014}.

In order to obtain physical results we need to fix the $G_v$ coupling constant.
 While in  QHD the non standard vector channel $\phi$ introduces
no new free parameters because they can be fixed throught symmetry
group arguments, unfortunately,  in the NJL,  this is not the case. In
most works the $G_v$ is treated  just as a free parameter
~\cite{DebPRC2009,DebPRC2014,MB,DebJCAP2019,Hana2001,Shao}.
Nevertheless,  in other works, the autors have tried to fix the
$G_v$ coupling  from direct comparisons with the lattice QCD
(LQCD) results. 

We follow this path.  In ref.~\cite{kit2002} studying the interplay
between chiral transition and CSC phase, the authors fixed
$G_V$ in the range 0.2 $<$ $G_v/G_s$ $<0.3$ in order to reproduce the
LQCD; in ref.~\cite{Contrera2014}  $G_v$ was fixed in the range 0.283 $<$ $G_v/G_s$ $<0.373$
in order to reproduce the slope of the pseudo-critical temperature for the chiral phase transition at low chemical potential 
extracted from LQCD simulations; also to reproduce the pseudo-critical
temperature, in ref.~\cite{Hell2011} $G_V$ was found to be in the range 0.25 $<$ $G_v/G_s$ $<0.4$ and finally, in ref.~\cite{Sugano2014}  a very restrictive choice was
made and the $G_v/G_s$ = 0.33. We follow the last suggestion and also
fix $G_v/G_s$ = 0.33 next.

\begin{figure*}[ht]
\begin{tabular}{cc}
\includegraphics[width=5.6cm,height=6.2cm,angle=270]{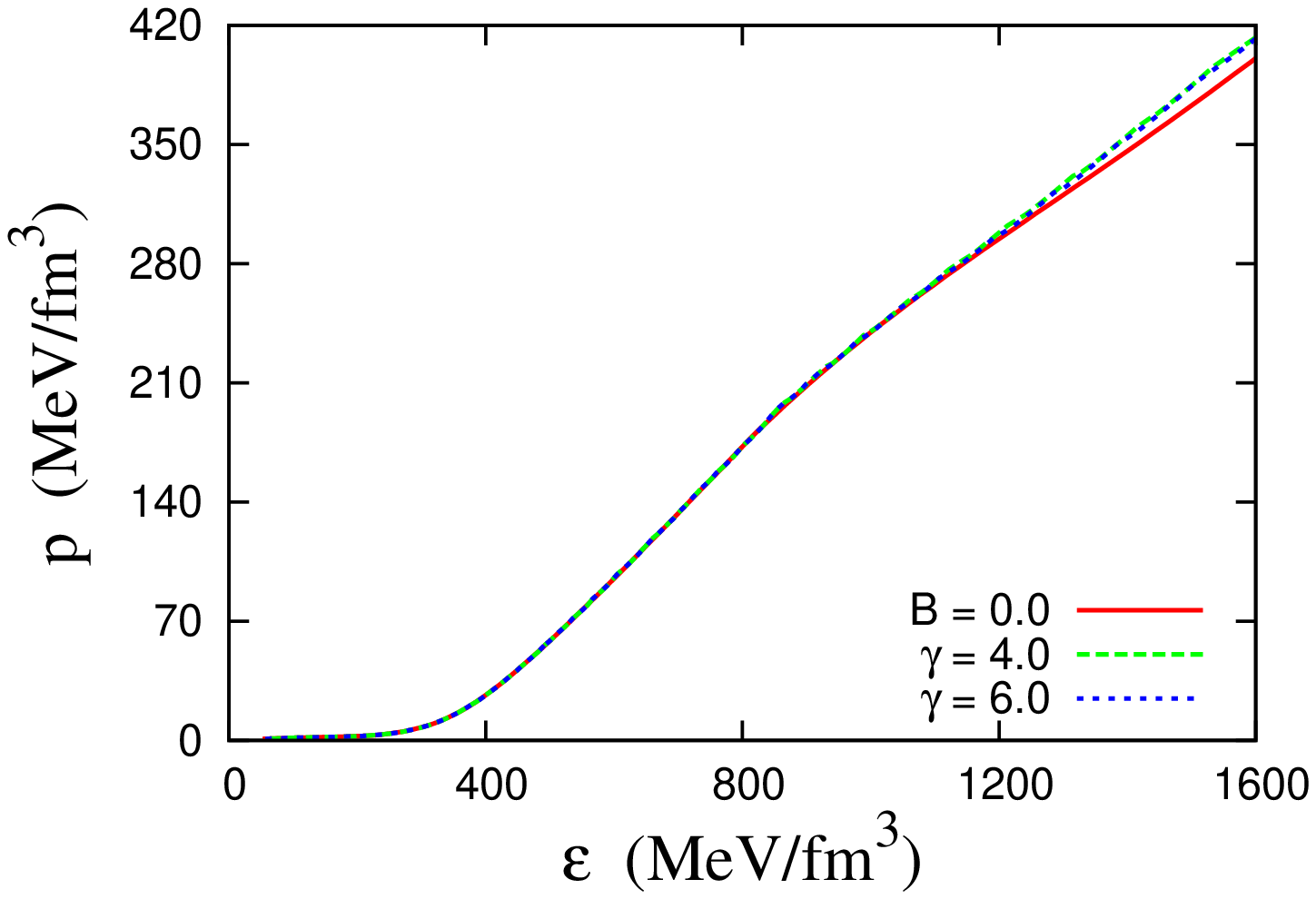} &
\includegraphics[width=5.6cm,height=6.2cm,angle=270]{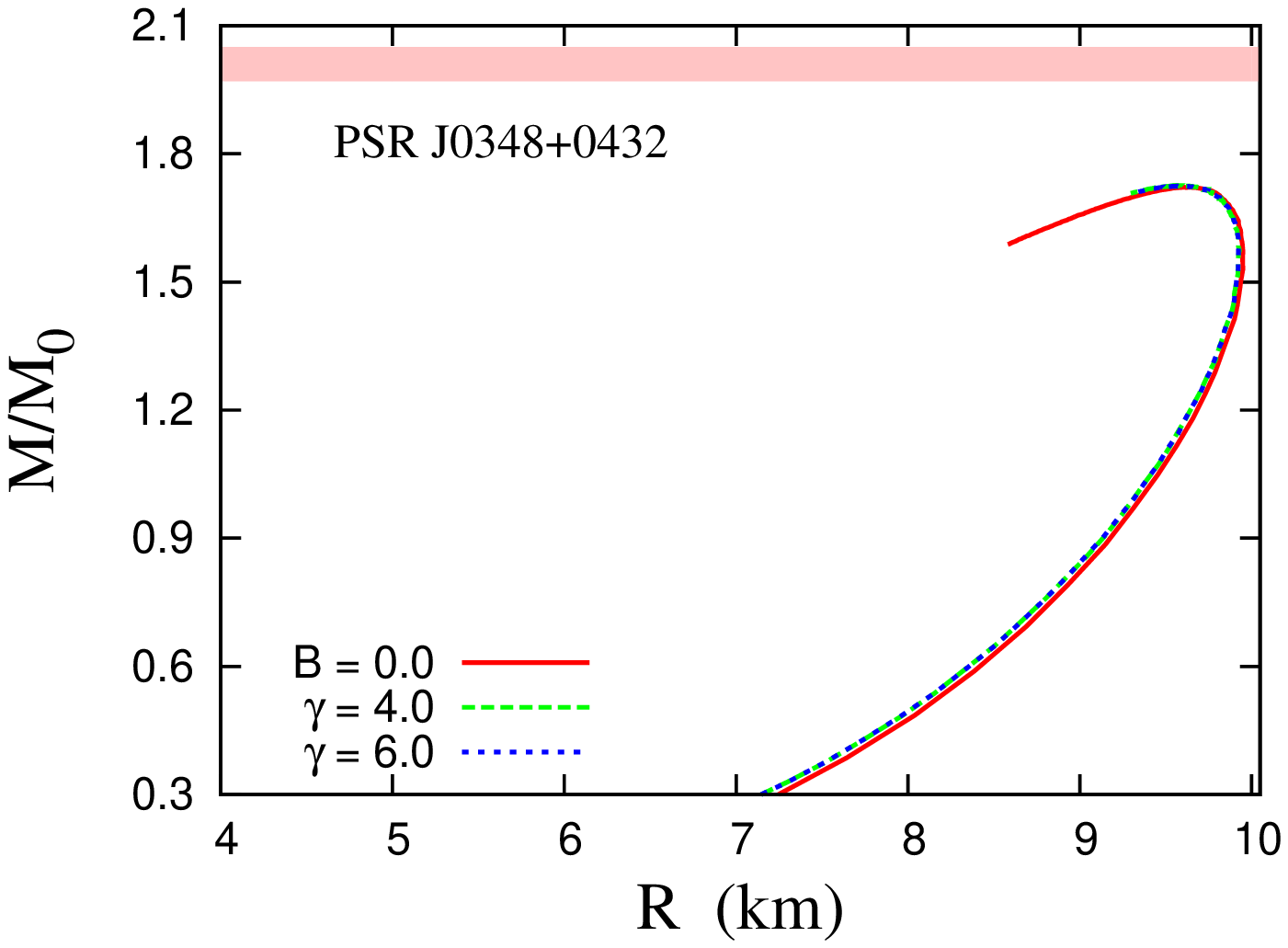} \\
\end{tabular}
\caption{(Color online) EoS and mass-radius relation for a quark
  star with zero magnetic field and two 
variations of magnetars with the $G_v$ channel.
No quark stars results  agree with the PSR J0348+0432. } \label{FL8}
\end{figure*}

The results for the particle population are exactly the same  as the ones obtained without the vector channel. This is 
due to the fact that all coupling constants are universal, coupling equally  the three quarks. This result
was already noted in ref.~\cite{DebPRC2014}. Due to this fact, we do not plot the fraction os particles here, as it can 
be seen in Fig.~(\ref{FL5}). 

In Fig.~(\ref{FL8}) we replot the EoS and the mass-radius relation
for quark matter and magnetars with the vector channel.  As the vector channel
enters in the pressure calculation, causing the EoS to become stiffer
we expect  higher maximum masses as we show  in Tab.~\ref{TL6}.
However, as we can see, the increase of the mass is not
enough to reproduce the PSR J0348+0432 pulsar. 
 In this work we use the parameters presented in ref.~\cite{Kun2}, which were obtained from the fitting of the physical quantities shown in Tab.~\ref{TL4}. With these physical quantities it is improbable that a NJL model be able to predict two solar masses quark stars, although this possibility is not completely ruled out. Another point worth mentioning is that we could have raised the maximum mass up to 2.0 solar masses by increasing the value of the $G_v$ coupling, as done in \cite{DebPRC2014}. However, instead of artificially increasing the vector channel, as it would violate LQCD results, we have chosen to face the consequences.  The
present study indicates that massive pulsars cannot be composed 
of deconfined quarks, at least if we believe that the NJL with the parametrization used in the present work is a reliable quark model. This does not completely rule out the possibility of quark stars.
As pointed in ref.~\cite{DebJCAP2019} and other references therein,
the hadronic neutron star could be a metastable system, which
  eventually collapses. If the original neutron star
had a mass beyond 1.74 $M_\odot$ it would become a black hole. 
However, a lower mass neutron star would become a quark star. The
  other possibility is that the metastable hadronic star can face a
  transition to a hybric star \cite{Melrose}, an object with both, hadron
  and quark matter \cite{hybrid}, as  discussed in the next section.

\begin{table}[ht]
\begin{tabular}{|c|c|c|c|c|c|}
\hline
 $\gamma$ & $M/M_\odot$ & $ R (km)$ & $\epsilon_c$ ($MeV/fm^3$) & $B_c$ ($ 10^{18}G)$ & $R_{1.4} (km)$   \\
\hline
 $B=0$ & 1.73 & 9.60   & 1521 & 0.0 & 9.89    \\
\hline
 4         & 1.74 & 9.57  &  1499 & 2.29 & 9.86   \\
 \hline
6          & 1.73 & 9.59  & 1478 & 2.10  & 9.86  \\
\hline
\end{tabular}
 \caption{Quark stars main properties with the $G_v$ coupling.
Even with this channel the maximum masses are not high enough to explain the massive PSR J0348+0432. }\label{TL6}
 \end{table}

The effect of the magnetic field only increases the mass by
0.01$M_\odot$, as all other models investigated with the chaotic
magnetic field approximation, i.e., the increase in the maximum mass
is very  subtle.

\section{Magnetars as Hybrid Stars}

As the density increases towards the star core, quarks become more
energetically favorable than baryons, and ultimately the
neutron star core may be composed of deconfined quarks. If the entire star
does not convert itself into a quark  star as suggested by the Bodimer-Witten conjecture, the final
composition is a quark core surrounded by a hadronic  layer. This is
what is generally called a hybrid star.
Nevertheless the nature of a hybrid star itself is not a closed
question. Some authors, as Hans Beth, and Brian Serot~\cite{Beth,Serot2}
claimed that the quark-hadron phase transition needs to be a
constant-pressure phase transition obtained through a Maxwell construction.
Within Maxwell constructions the quark and the hadron phases  are
spatially  separated and there is a discontinuity in the electron
chemical potential, although the nucleon chemical potentials are continuous.
On the other hand Glenndening ~\cite{Glen} suggested that Maxwell
construction is $\beta$ unstable at the interface between the
phases and proposed a construction in which the  pressure and all chemical
  potentials (including the electron chemical potential) are
  continuous, the one called Gibbs construction. Under these construction
quarks and hadrons coexist in a mixed state, generating one
  intermediate phase in between the hadronic and the quark phases.
 Despite these
differences,  some authors ~\cite{Chiba1,Chiba2,Paoli} performed studies on hybrids stars
with both, Maxwell and Gibbs construction. They all concluded that
there is no significant difference on the macroscopic properties of the
hybrid stars. Due to this fact, in this work we use a Maxwell
construction, which is simpler. In this case we impose that the transition 
occurs when the pressure of the quark matter equals the pressure of the hadronic matter at the same chemical potential:

\begin{equation}
\mu^H_n = \mu^Q_n \quad \mbox{and} \quad p^H = p^Q, \label{EMax}
\end{equation}
where the neutron chemical potential can be written in terms of the quark ones as:
\begin{eqnarray}
\mu_d = \mu_s = \frac{1}{3} ( \mu_n + \mu_e) \nonumber \\
\mu_u = \frac{1}{3} (\mu_n - 2\mu_e) .
\end{eqnarray}

\begin{figure*}[ht]
\begin{tabular}{cc}
\includegraphics[width=5.6cm,height=6.2cm,angle=270]{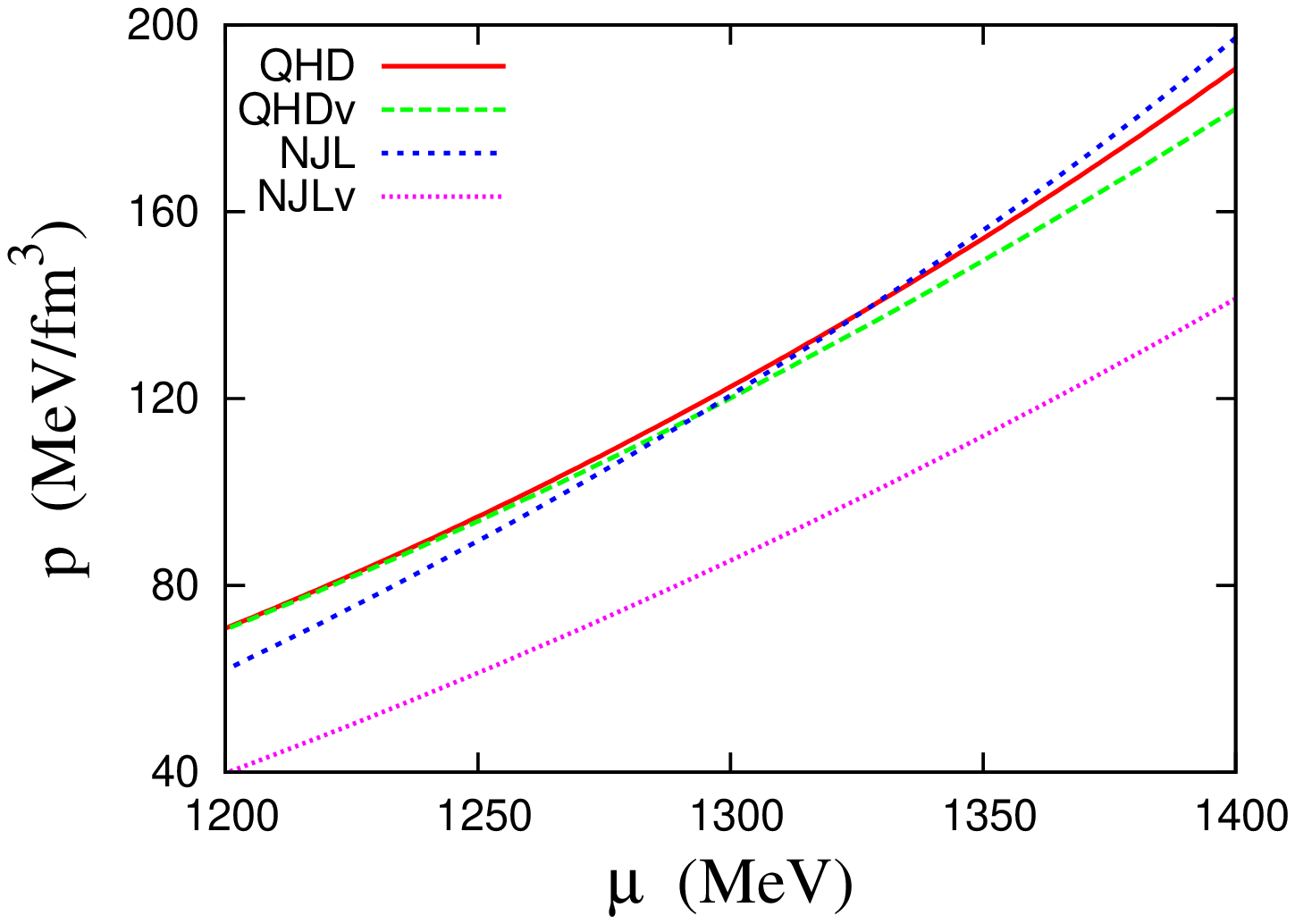} &
\includegraphics[width=5.6cm,height=6.2cm,angle=270]{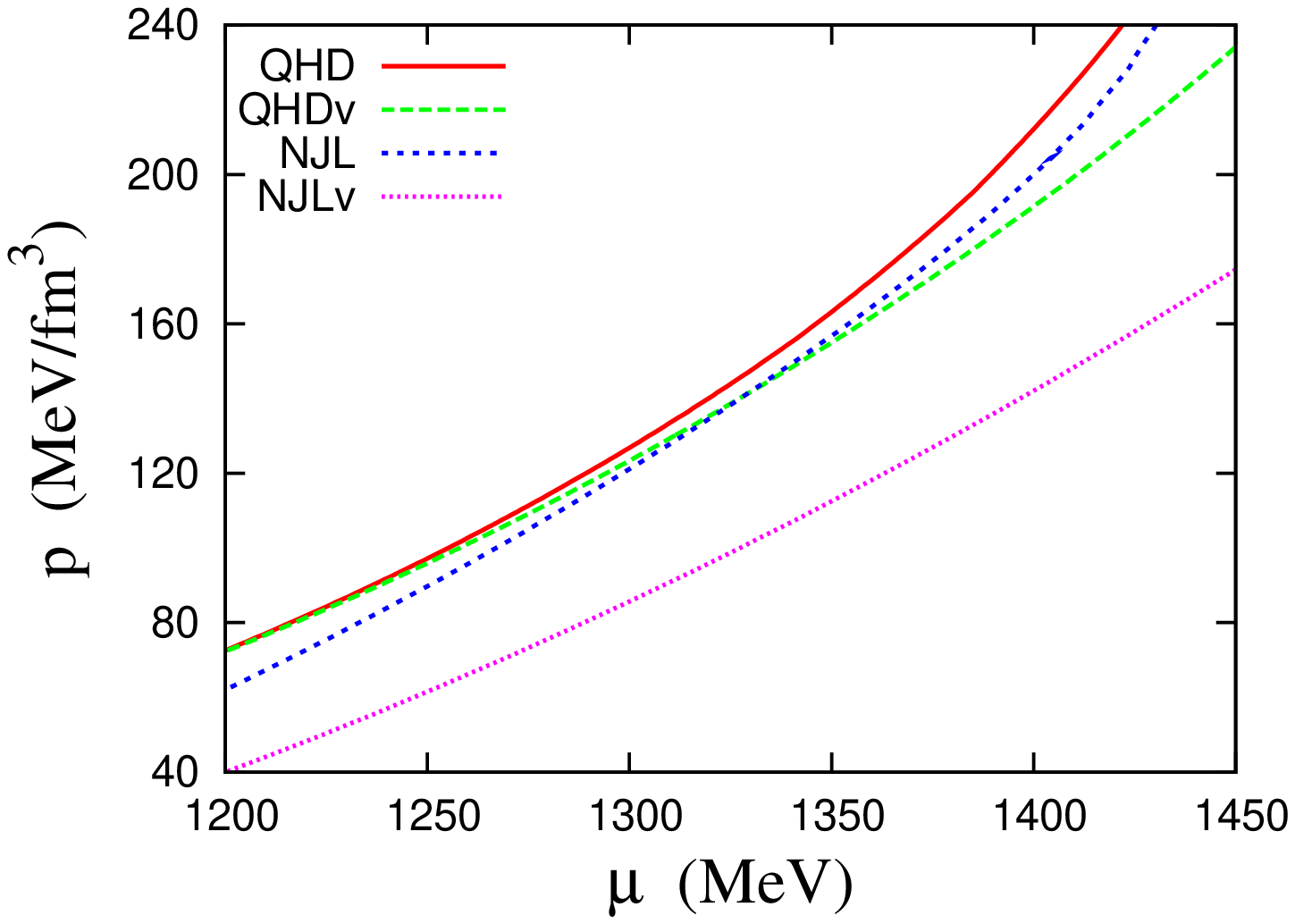} \\
\end{tabular}
\caption{(Color online)  Pressure as function of the chemical potential
  with (right) and without (left) magnetic field (B= $3 \times 10^{18}$ G). If a hadronic EoS crosses a quark
  one, the phase with deconfined quarks becomes energetically
  favorable at higher densities.} \label{FL9}
\end{figure*}

We have previoulsy presented two possibilities to describe hadronic
matter, with and without the vector channel (let's call them QHD and
QHDv respectively). We have also shown two
ways to describe quark matter (NJL and NJLv). Moreover each
possibility has been studied in two configurations of the magnetic field. Now we can study how
each of these factor influences the formation of hybrid stars. To accomplish that, we plot in Fig.~(\ref{FL9}) the pressure as function of the chemical
potential, and seek for the point that satisfies the conditions presented in Eq.~(\ref{EMax}) with and without magnetic field.

We see that the vector channel in the hadron phase increases the pressure, so it induces the hadron quark phase transition at early densities when
compared with the case without the $\phi$ meson. On other hand, the
vector channel in the quark phase increases the pressure as well.
 With the inclusion of these two vector channels, the pressure becomes
higher in the quark phase than in the hadronic one,  thus suppressing
the  phase transition. Indeed, if we believe that the NJLv model is a good model to describe
quark matter, therefore it is not probable that hybrid stars exist
 since, as we can see, Eq.~(\ref{EMax}) is not satisfied. 

More interesting phenomena, however, appear  when we deal with the
magnetic field. As pointed  out during our discussions on
hadronic and quark stars, the effect of the magnetic field on the EoS is not relevant,
causing only a small increase on the maximum masses. However, when we
seek for both phases crossing point we see that the magnetic field
plays the role of suppressing the quark onset. 
As we can see from  the study involving QHDv to NJL the chemical potential at the phase
transition  point is pushed away from 1295 $MeV$ to 1340 $MeV$.
 When QHD and NJL models are considered, 
the chemical potential moves from 1330 $MeV$ to 1439
$MeV$. Note that such high chemical potential is already above the
  one corresponding to the maximum mass of a hadronic neutron star. 
The fact that the magnetic field strongly difficults the formation of
a hybrid star was already pointed out in the
recent literature~\cite{Dex4,Dex5}. Our study corroborates these
results.
  The phase transition pressure and chemical potential values are presented in Tab.~\ref{TL7}.

\begin{widetext}
\begin{center}
\begin{table}[ht]
\begin{center}
\begin{tabular}{|c|c||c|c|c|c|c|c|}
\hline 
 Hadron & Quark & B ($\times 10^{18}$) G & $\mu_n^H = \mu_n^Q$  & $p^H = p^Q$ & $\epsilon^H$ $(MeV/fm^3)$ & $\epsilon^Q$ $(MeV/fm^3)$ \\
 \hline
 QHD & NJL  & 0.0 & 1330 $MeV$ & 141 $(MeV/fm^3)$ & 711 & 808 \\
 \hline
 QHDv & NJL  & 0.0 & 1295 $MeV$ & 117 $(MeV/fm^3)$ & 597 & 730 \\
 \hline
 QHD & NJL  & 3.0 & 1439 $MeV$  & 264 $(MeV/fm^3)$ & 1130 & 1409 \\
  \hline
 QHDv & NJL  & 3.0 & 1340 $MeV$  & 149 $(MeV/fm^3)$ & 683  & 834 \\
  \hline
\end{tabular}
\caption{Chemical potential and pressure at phase transition for QHD to NJL with and without the vector channel and the magnetic field.
We also present the energy density of each matter at the phase transition. The NJLv model does not allow a phase transition.}
\label{TL7}
\end{center}
\end{table}
\end{center}
\end{widetext}

\begin{figure*}[ht]
\begin{tabular}{cc}
\includegraphics[width=5.6cm,height=6.2cm,angle=270]{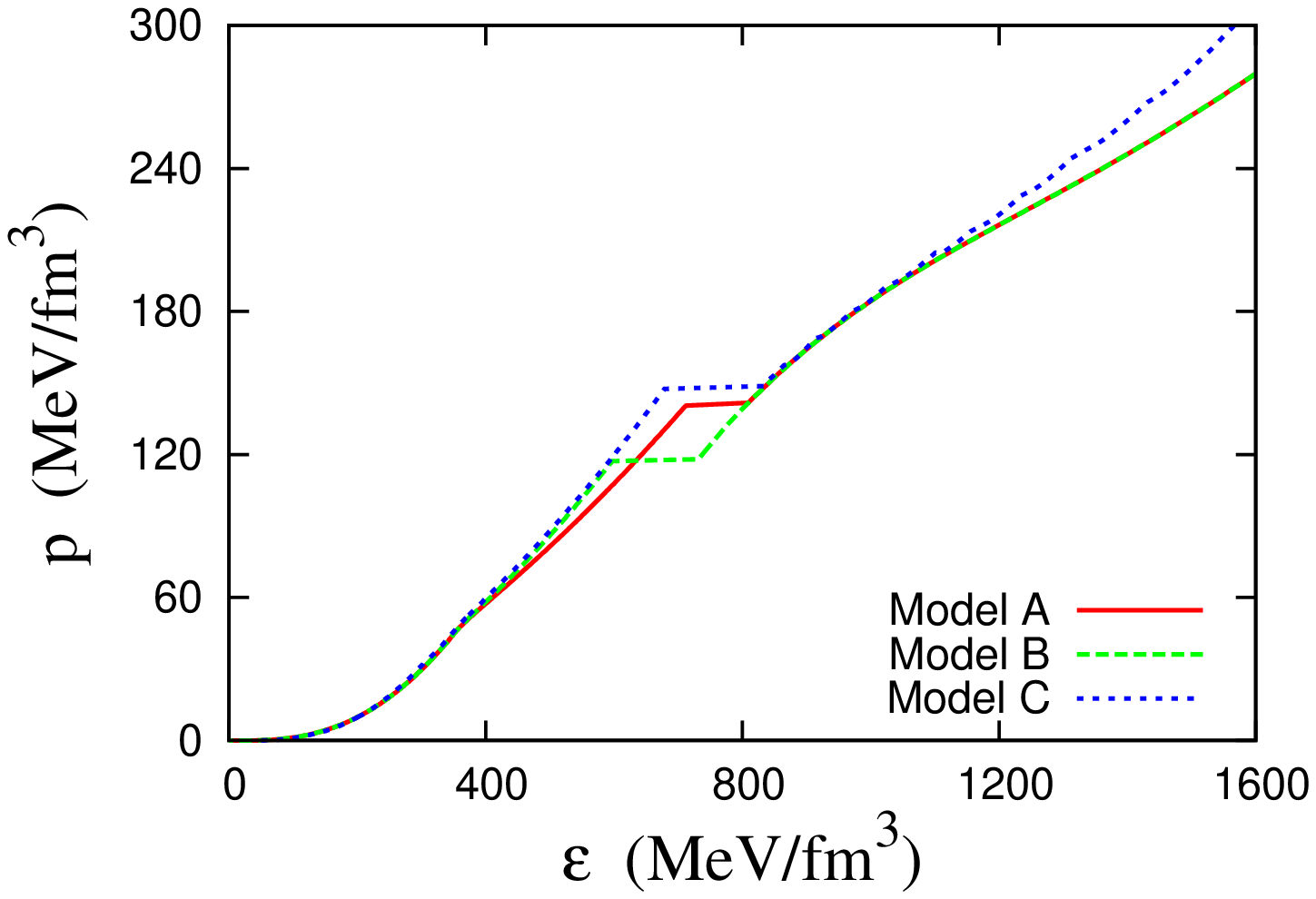} &
\includegraphics[width=5.6cm,height=6.2cm,angle=270]{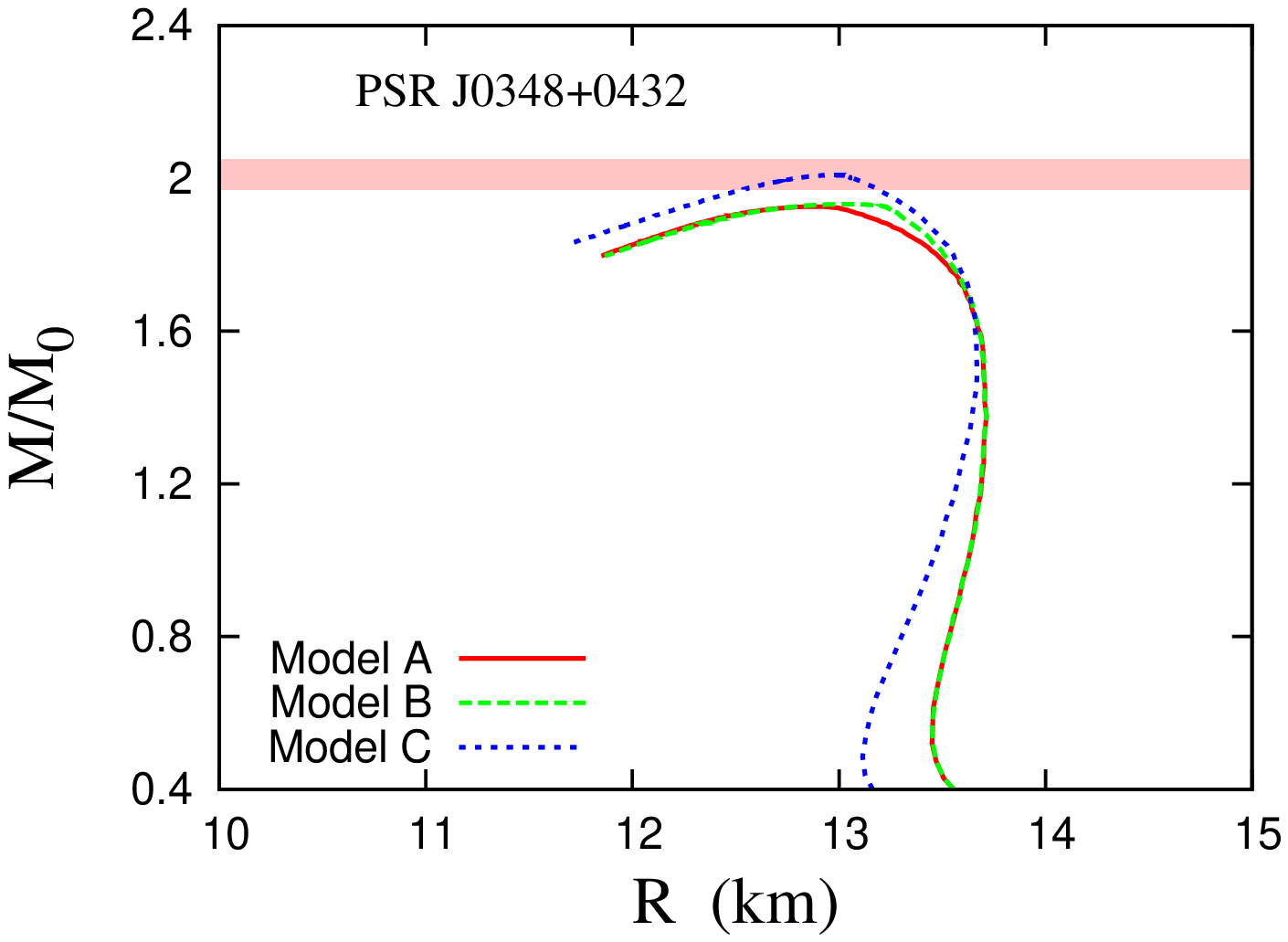} \\
\end{tabular}
\caption{(Color online) EoS and mass-radius relation for models that allow a construction of an hybrid star.} \label{FL10}
\end{figure*}

As can be seen, there are three possible models of hybrid stars that
we now nominate: model A corresponds to the EoS obtained from
QHD and NJL without magnetic field, model B  to QHDv and NJL also with no
  magnetic field, and finally model C to the QHDv to NJL with the
  inclusion of magnetic field.  We plot the EoS and the mass-radius
  diagram for them in Fig.~(\ref{FL10}) and resume the main properties
  in Tab.~\ref{TL8}. 

\begin{table}[ht]
\begin{tabular}{|c|c|c|c||c|c|}
\hline
 Model & $M/M_\odot$ & $ R (km)$ & $\epsilon_c$ ($MeV/fm^3$) & $M_{min}/M_\odot$ & $R_{1.4} (km)$   \\
\hline
 A & 1.93  & 12.88   & 877   & 1.91 & 13.63  \\
\hline
 B         & 1.94  & 13.04  &  822   & 1.92 & 13.63 \\
 \hline
C        & 2.02 & 12.97  & 879   & 2.00 & 13.58 \\
\hline
\end{tabular}
 \caption{Hybrid star properties indicating that  the allowed mass
   values for a quark core lie within a very narrow band. }\label{TL8}
 \end{table}
 
The maximum mass of a hybrid neutron star without magnetic field 
  lies just slightly below the lower limit of the PSR J0348+0432.
When the magnetic field is  included, we see that the  PSR J0348+0432
mass constraint is satisfied. Notice that hybrid stars  bear the lower
value of the energy density at the core as  compared with hadronic
  and quark stars, and thus lower values of the magnetic fields. 
Also they have the bigger radii for the maximum masses
when compared with hadronic and quark stars. We also investigate the
$M_{min}$, which is the  minimum mass  star that supports a quark core, i.e., 
stars with masses below  $M_{min}$ are purely hadronic. We see that,
although the existence of a hybrid star is possible, it is very
unlikely, because there is just a narrow mass range that 
  supports a quark core. Indeed if we move away from the maximum mass by only 0.02$M_\odot$
we obtain a pure hadronic star, instead of a hybrid one. This small
range is even below the experimental  uncertainty on the mass of the
PSR J0348+0432. Hence, its true nature is still an open subject,
although  probably it is not a quark star.

\section{Final Remarks}

In this work we studied the influence of a non standard repulsive
channel and strong magnetic field in different classes of 
compact stars. The results can be summarized:

\begin{itemize}

\item The $\phi$ meson in the hadronic matter significantly increases the pressure, making hadronic neutron stars fully compatible with the PSR J0348+0432,  and even with the more massive MSP J0740+6620. As this meson does not affect nuclear properties and brings no new parameters, since we can constrain $g_\phi$ with the help of symmetry group, its presence seems to improve the description of hadronic matter at high densities.

\item  The $G_v$ in the quark matter also increases the pressure. However there are two problems: First, unlike the hadronic case, 
this new vector channel cannot be fixed by theory, only by
phenomenology. Second, although there is some consensus 
about the possible values of $G_v$ from lattice QCD, even with this term the quark stars are not massive enough to describe the  PSR J0348+0432.
 An arbitrary increase of $G_v$ in order to accomplish that seems highly unphysical.
 
 \item { As far as the chaotic magnetic field is concerned, some points are worth emphasizing: $1-)$ In our previous work~\cite{Lopes2015}, we compared the effect of the chaotic magnetic field with those obtained by Lorene available at that date and obtained very similar results. $2-)$ The chaotic magnetic field restores the thermodynamic pressure. $3-)$ The chaotic magnetic field does not overestimate the neutron star mass as stated in previous works and also reproduced here. }

\item The magnetic field has very little influence on the macroscopic   properties of hadron and quark stars,  but its inclusion makes the formation of hybrid stars more difficult, pushing away the  critical neutron chemical potential. 

\item The additional vector channel in QHD favors the existence of   hybrid stars, while in NJL it suppresses this possibility  completely.

\item  According to the model presented in this work, the minimum and the maximum masses of  a hybrid star are very   close to each other,  indicating that only stars at the edge of mechanical stability can be  hybrid stars. If they exist, hybrid stars are very rare in the universe,  although the results are model dependent.

\item Quark stars can still be present in nature, if the Bodmer-Witten
  conjecture is true.  Low masses hadronic neutron stars can
collapse to form a quark star. Massive neutron stars collapse into black holes. 

\item We can describe even more massive neutron stars if hyperons are not
  present. 
The hyperon puzzle is still open and can only be accounted for 
  under special circumstances.

\end{itemize}

 Moreover, it is worth noticing that the recent detection of gravitational waves from neutron stars merger has opened a new window to constrain dense nuclear matter \cite{LIGO}. A clear correlation between dimensionless tidal deformability and the radius of the canonical star was found (see, for instance,\cite{malik,Odilon1}). Some studies~\cite{zhao,elias} constrain the radii of the canonical stars to 12.00 km $<~R_{1.4}~<$ 13.45 km, while in \cite{malik}, the proposed range is  $11.82~\mbox{km} \leqslant R_{1.4}\leqslant  13.72~\mbox{km}$ and in \cite{Odilon2} it is $10.36~\mbox{km} \leqslant R_{1.4}\leqslant  12.55~\mbox{km}$.
By using the neutron skin values as a new 
constraint, the upper limit for $R_{1.4}$ was set to be $13.76~\mbox{km}$~\cite{fatt18}. There is no direct evidence that the neutron stars in the binary system are magnetars, but anyway,
our results in this work (around 13.6 km) are compatible with many of the constraints just mentioned.

Before finishing this work, as pointed out in the introduction, we discuss in more detail our
results and the validity of the chaotic magnetic field approximation  
in the light of a new study published in ref.~\cite{debi2018}
 concerning the magnetic field distribution in magnetar interiors. 
Using full relativistic numerical calculation, it was found
that the magnetic field can be expressed as a multipolar expansion that
accounts for the monopole contribution, the dipole term, the quadrupole
and so on. Now, the important fact here is is that our study is
based on the chaotic magnetic field
formalism, which is a monopole approximation for the magnetic field
profile. However as shown in Fig 3 of \cite{debi2018} the monopole
term is dominant in almost the entire star. 
More than that, the monopole term is specially dominant in the neutron star
core, when the magnetic field is stronger.  So, in the limit of very
high field,  when its influence is bigger, our results are very close to those obtained in this work. Moreover, one of the main problems of using
TOV equations in the presence of strong magnetic fields is the
possible appearance of anisotropies in the momentum-energy tensor.
As pointed out in ref.~\cite{debi2018} in most cases, $T^{\theta\theta} ~ \neq ~ T^{rr}$. 
However, exactly due to the monopole nature of the chaotic magnetic
field, we always obtain $T^{\theta\theta} = T^{rr}$,
which guarantees that the TOV approximation is a good one in this
case. 

{\bf Acknowledgments} 

This work is a part of the project INCT-FNA Proc. No. 464898/2014-5 and it was partially supported by CNPq (Brazil)
under grant 301155.2017-8 (D.P.M).

\end{document}